\documentclass[aps,prb,twocolumn,superscriptaddress,showpacs]{revtex4-1}
\def\be{\begin{equation}}
\def\ee{\end{equation}}
\def\bea{\begin{eqnarray}}          
\def\eea{\end{eqnarray}}
\def\bi{\begin{itemize}}
\def\ei{\end{itemize}}

\usepackage{graphicx}

\begin{document}

\title{ 
              Projected Entangled Pair States at Finite Temperature:  \\
   Iterative Self-Consistent Bond Renormalization for Exact Imaginary Time Evolution
}

\author{Piotr Czarnik} 
\affiliation{Instytut Fizyki Uniwersytetu Jagiello\'nskiego,
             ul. {\L}ojasiewicza 11, 30-348 Krak\'ow, Poland}

\author{Jacek Dziarmaga} 
\affiliation{Instytut Fizyki Uniwersytetu Jagiello\'nskiego,
             ul. {\L}ojasiewicza 11, 30-348 Krak\'ow, Poland}

\date{ November 23, 2014 }

\begin{abstract}
A projected entangled pair state (PEPS) with ancillas can be evolved in imaginary time 
to obtain thermal states of a strongly correlated quantum system on a 2D lattice. 
Every application of a Suzuki-Trotter gate multiplies the PEPS bond dimension $D$ by a factor $k$.
It has to be renormalized back to the original $D$. 
In order to preserve the accuracy of the Suzuki-Trotter (S-T) decomposition,
the renormalization has in principle to take into account full environment made of 
the new tensors with the bond dimension $k\times D$.
Here we propose a self-consistent renormalization procedure operating with the original bond dimension $D$,
but without compromising the accuracy of the S-T decomposition.
The iterative procedure renormalizes the bond using 
full environment made of renormalized tensors with the bond dimension $D$.
After every renormalization, 
the new renormalized tensors are used to update the environment, 
and then the renormalization is repeated again and again until convergence.
As a benchmark application,
we obtain thermal states of the transverse field quantum Ising model on a square lattice -
both infinite and finite -
evolving the system across a second-order phase transition at finite temperature.
\end{abstract}
\maketitle

%%%%%%%%%%%%%%%%%%%%%%%%%%%%%%%%%%%%%%%%%%%%%%%%%%%%%%%%%%%%%%%%%%%%%%%%%% 
\section{ Introduction } 
%%%%%%%%%%%%%%%%%%%%%%%%%%%%%%%%%%%%%%%%%%%%%%%%%%%%%%%%%%%%%%%%%%%%%%%%%% 

Quantum tensor networks are a competitive tool to study strongly correlated quantum systems on a lattice. 
Their history begins with the density matrix renormalization group (DMRG) \cite{White} - 
an algorithm to minimize the energy of a matrix product state (MPS) ansatz in one dimension (1D), 
see Ref. \cite{Schollwoeck} for a review of MPS algorithms. 
In the last decade, 
MPS was generalized to a 2D ``tensor product state'' widely known as a projected entangled pair state (PEPS) \cite{PEPS}. 
Another type of tensor network is the multiscale entanglement renormalization ansatz (MERA) \cite{MERA}, 
and the branching MERA \cite{branching}, 
that is a refined version of the real space renormalization group. 
Being variational methods, 
the quantum tensor networks do not suffer form the notorious fermionic sign problem, 
and thus they can be applied to strongly correlated fermions in 2D \cite{fermions}. 
A possible breakthrough in this direction is an application of the PEPS ansatz to the t-J model \cite{PEPStJ}, 
which is a strong coupling approximation to the celebrated Hubbard Hamiltonian of the high temperature superconductivity \cite{highTc}. 
An energy of the ground state was obtained that could compete with the best variational Monte-Carlo results \cite{VMC}.

The tensor networks also proved to be a powerful tool to study topological spin liquids (TSL). The search for 
realistic models gained momentum after White demonstrated the spin-liquid nature of the Kagome antiferromagnet 
\cite{WhiteKagome}. This result was obtained by a {\it tour de force} application of a quasi-1D DMRG. The DMRG 
investigation of TSL's was elevated to a higher degree of sophistication in Ref. \cite{CincioVidal}. Unfortunately, 
the MPS tensor network underlying the DMRG suffers from severe limitations in two dimensions, where it can be used 
for states with a very short correlation length only. In contrast, the PEPS ansatz in Fig. \ref{FigPeps} is not 
restricted in this way. Its usefulness for TSL has already been demonstrated. In Ref. \cite{PepsRVB} it was shown 
how to represent the RVB state with the PEPS ansatz in an efficient way. In Ref. \cite{PepsKagome} PEPS was used 
to classify topologically distinct ground states of the Kagome antiferromagnet. Finally, in Ref. \cite{PepsJ1J2} 
PEPS demonstrated a TSL in the antiferromagnetic $J_1-J_2$ model.

In contrast to the ground state, finite temperature states have been explored so far mostly with the MPS 
\cite{ancillas,WhiteT}. In a way that can be easily generalized to 2D, the MPS is extended to finite 
temperature by appending each lattice site with an ancilla \cite{ancillas}. A thermal state is obtained by 
an imaginary time evolution of a pure state in the enlarged Hilbert space starting from infinite temperature. 
However, thermal states are of more interest in 2D, where they can undergo finite temperature phase transitions.  
A thermal PEPS with ancillas was considered in Ref. \cite{Czarnik}, where finite temperature states
of the 2D quantum Ising model and a spinless fermionic system were obtained. 
Alternative approaches to finite temperature were developed \cite{ChinaT} where, 
instead of the imaginary time evolution, 
a tensor network representing the partition function is directly contracted by subsequent tensor renormalizations. 
Another interesting alternative is based on linear optimization of local density matrices at finite $T$ \cite{Poulin}. 

Here we revisit the approach of Ref. \cite{Czarnik} with the aim to improve its numerical efficiency.
After every infinitesimal time step effected by a Suzuki-Trotter (S-T) gate, 
the bond dimension $D$ of a PEPS tensor is multiplied by a factor $k\geq2$. 
This dimension has to be truncated/renormalized back to the original $D$ in a way that distorts the new PEPS the least. 
To preserve the accuracy of the S-T decomposition, 
the renormalization has to take into account full environment of the renormalized bond and 
after the gate the environment is made of tensors with the enlarged bond dimension $k\times D$.
The infinite environment is calculated with the help of the corner matrix renormalization \cite{CMR}.
It is the most time-consuming part of the time-evolution algorithm that needs to be accelerated.
In this paper, 
we propose a self-consistent renormalization scheme that is using 
full environment made of renormalized tensors with the original bond dimension $D$. 
After every renormalization, the new renormalized tensors are used to update
the environment and then the renormalization is repeated again and again until convergence.
The converged renormalized tensors are accepted as the new PEPS tensors after the S-T gate.
As a benchmark application,
we evolve the quantum Ising model on a square lattice - 
both infinite and finite - 
across a finite-temperature second-order phase transition in a strong transverse magnetic field.

The paper is organized as follows. In Section \ref{SecPurification} we introduce purifications of 
thermal states represented by PEPS and in Sections \ref{SecIsing} and \ref{SecST} we remind the reader 
of the quantum Ising model and the Suzuki-Trotter decomposition in the PEPS formalism respectively. 
Section \ref{SecW} explains how the enlarged bond dimension can be truncated back to the original 
size $D$ with the help of an isometry. The iterative self-consistent optimization of the isometry 
is introduced in Section \ref{SecSelf} that is supplemented by Appendices \ref{CMR} and \ref{SecAcc}.
The benchmark results in the Ising model on an infinite lattice are presented in Section \ref{SecBenchmark} -
supplemented by Appendix \ref{SecOnsager} - and on a finite one in Sec. \ref{SecFinite}.
Finally, we conclude in Section \ref{SecConclusion}.

%%%%%%%%%%%%%%%%%%%%%%%%%%%%%%%%%%%%%%%%%%%%%%%%%%%%%%%%%%%%%%%%%%%%%%%%%% 
\section{ Purification of thermal states as PEPS\label{SecPurification}} 
%%%%%%%%%%%%%%%%%%%%%%%%%%%%%%%%%%%%%%%%%%%%%%%%%%%%%%%%%%%%%%%%%%%%%%%%%%

We consider spins on an infinite square lattice with a Hamiltonian ${\cal H}$. 
Every spin has $S$ states $i=0,...,S-1$ and is accompanied by an ancilla with states $a=0,...,S-1$. 
The enlarged Hilbert space is spanned by states $\prod_m |i_m,a_m\rangle$, 
where the product runs over lattice sites $m$. 
The state of spins at infinite temperature,
$
\rho(\beta=0) = \prod_m \left( \frac{1}{S} \sum_{i=0}^{S-1} | i_m \rangle\langle i_m| \right) \propto {\bf 1},
$
is obtained from its purification in the enlarged Hilbert space, 
\be 
\rho(0) ~=~ {\rm Tr}_{\rm ancillas}|\psi(0)\rangle\langle\psi(0)|~,
\ee
where 
\be 
|\psi(0)\rangle ~=~ \prod_m \left( \sum_{i=0}^{S-1} \frac{1}{\sqrt{S}} |i_m,a_m\rangle \right)~
\label{psi0}
\ee
is a product of maximally entangled states of every spin with its ancilla. The state $\rho(\beta)\propto e^{-\beta {\cal H}}$ 
at finite $\beta$ is obtained from the purification
\be 
|\psi(\beta)\rangle~\propto~
e^{-\frac12\beta {\cal H}}~|\psi(0)\rangle~\equiv~
U(\beta)~|\psi(0)\rangle~
\ee
after imaginary time evolution for time $\beta$ with $\frac12{\cal H}$.

\begin{figure}[h!]
%\vspace{-0.5cm}
\includegraphics[width=1.0\columnwidth,clip=true]{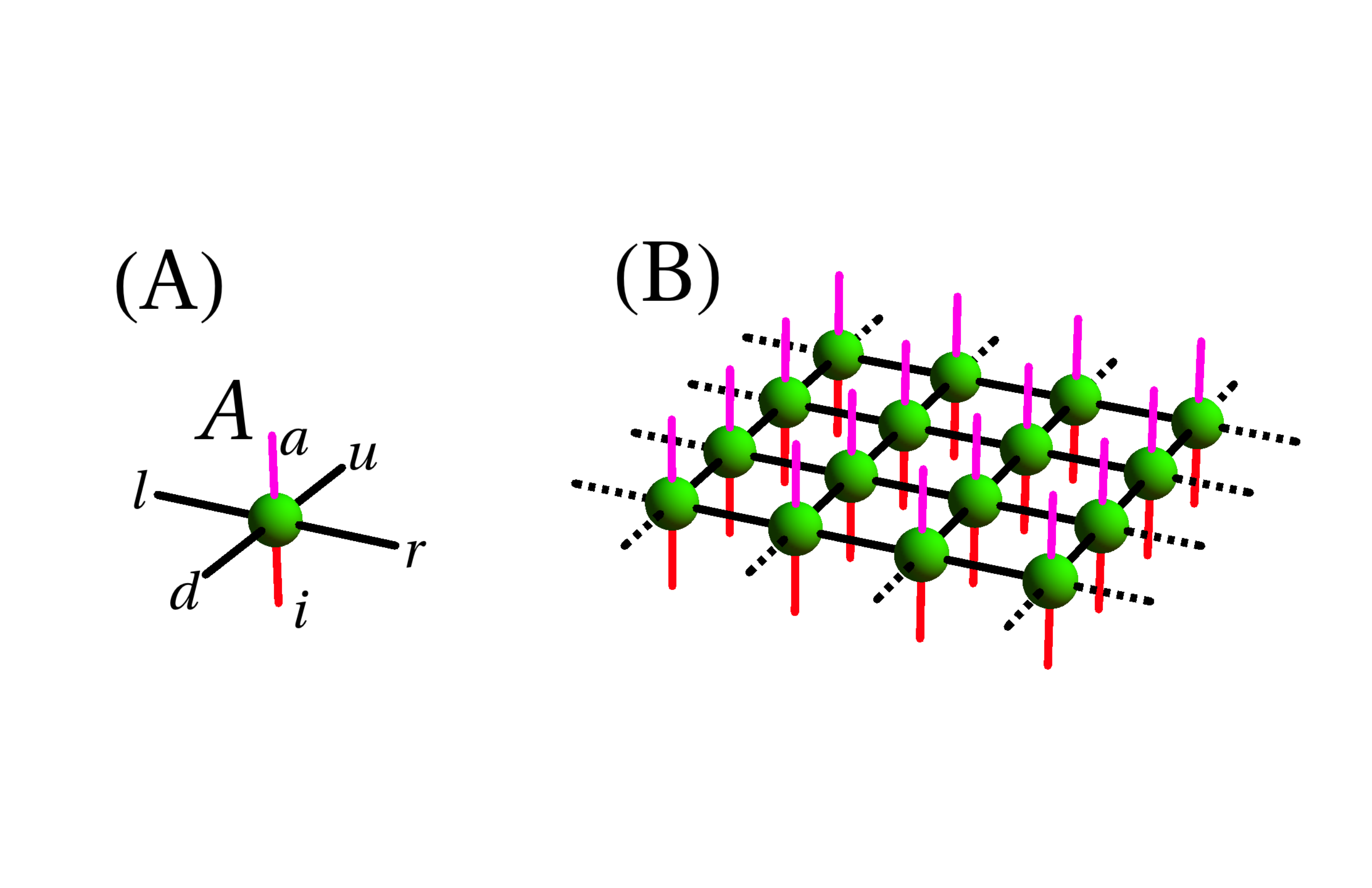}
\vspace{-1.5cm}
\caption{ 
In A, 
graphic representation of the tensor $A^{ia}_{urdl}$.
In B, 
the amplitude $\Psi_A[\{i,a\}]$ with all bond indices connecting nearest-neighbor tensors contracted. 
Here and in the following figures the index contraction is represented by a line connecting two tensors.
}
\label{FigPeps}
\end{figure}

For an efficient simulation of the time evolution, 
we represent $|\psi(\beta)\rangle$ by a translationally invariant PEPS with the same tensor $A^{ia}_{trbl}(\beta)$ at every site. 
In the quantum Ising model that we consider in this paper, 
translational invariance is not broken and a unit cell encloses one lattice site. 
Here $i$ and $a$ are the spin and ancilla indices respectively, 
and $u,r,d,l=0,...,D-1$ are bond indices to contract the tensor with similar tensors at the nearest neighbor sites, 
see Fig. \ref{FigPeps}A. 
The ansatz is
\be 
|\psi(\beta)\rangle ~=~  \sum_{\{i_m,a_m\}} ~ \Psi_A[\{i_m,a_m\}] ~ \prod_m|i_m,a_m\rangle~ \equiv~ \left|\psi_A\right\rangle.
\ee
Here the sum runs over all pairs of indices $i_m,a_m$ at all sites $m$. 
The amplitude $\Psi_A$ is the tensor contraction in Fig. \ref{FigPeps}B. 
The initial state (\ref{psi0}) can be represented by a tensor 
\be 
A^{ia}_{urdl}~=~ \delta^{ia} ~ \delta_{u0} ~ \delta_{r0} ~ \delta_{d0} ~ \delta_{l0}~
\label{A0}
\ee
with the minimal bond dimension $D=1$.

%%%%%%%%%%%%%%%%%%%%%%%%%%%%%%%%%%%%%%%%%%%%%%%%%%%%%%%%%%%%%%%%%%%%%%%%%%%%
%%%%%%%%%%%%%%%%%%%%%%%%%%%%%%%%%%%%%%%%%%%%%%%%%%%%%%%%%%%%%%%%%%%%%%%%%%%%

%%%%%%%%%%%%%%%%%%%%%%%%%%%%%%%%%%%%%%%%%%%%%%%%%%%%%%%%%%%%%%%%%%%%%%%%%% 
\section{ Transverse field Quantum Ising model on a square lattice\label{SecIsing} } 

We proceed with
\be 
{\cal H} ~=~- 
\sum_{\langle m,m'\rangle}Z_mZ_{m'}
-h \sum_m X_m 
~\equiv~ 
{\cal H}_{ZZ}+{\cal H}_X.
\label{calH}
\ee
Here $Z,X$ are Pauli matrices. 
The model has a ferromagnetic phase 
with a non-zero spontaneous magnetization $\langle Z \rangle$ for 
small $h$ and large $\beta$. 
At $h=0$ the critical point is $\beta_0=-\ln(\sqrt{2}-1)/2=0.441$ 
and at zero temperature the quantum critical point is $h_0=3.044$, see Ref. \cite{hc}.

%%%%%%%%%%%%%%%%%%%%%%%%%%%%%%%%%%%%%%%%%%%%%%%%%%%%%%%%%%%%%%%%%%%%%%%%%% 
\section{ Suzuki-Trotter decomposition\label{SecST} }
  
We define 
$ 
U_{ZZ}(\Delta\beta) \equiv e^{-\frac12{\cal H}_{ZZ}\Delta\beta}
$ 
and 
$
U_X(\Delta\beta) \equiv e^{-\frac12{\cal H}_X\Delta\beta}
$ 
for the interaction and the transverse field respectively. 
In the second-order Suzuki-Trotter decomposition a small time step is a product
\be
U(d\beta) ~=~ U_X(d\beta/2)U_{ZZ}(d\beta)U_X(d\beta/2)~+~{\cal O}(d\beta^3).
\ee
The action of $U_X(d\beta/2)$ on PEPS replaces $A^{ia}_{urdl}$ with
\be 
\cosh\frac{hd\beta}{4}~
A^{ia}_{urdl}~+~
\sinh\frac{hd\beta}{4}~
\sum_j X^{ij} A^{ja}_{urdl} 
\ee
of the same bond dimension $D$. 

\begin{figure}[h!]
%\vspace{-0.5cm}
\includegraphics[width=1.0\columnwidth,clip=true]{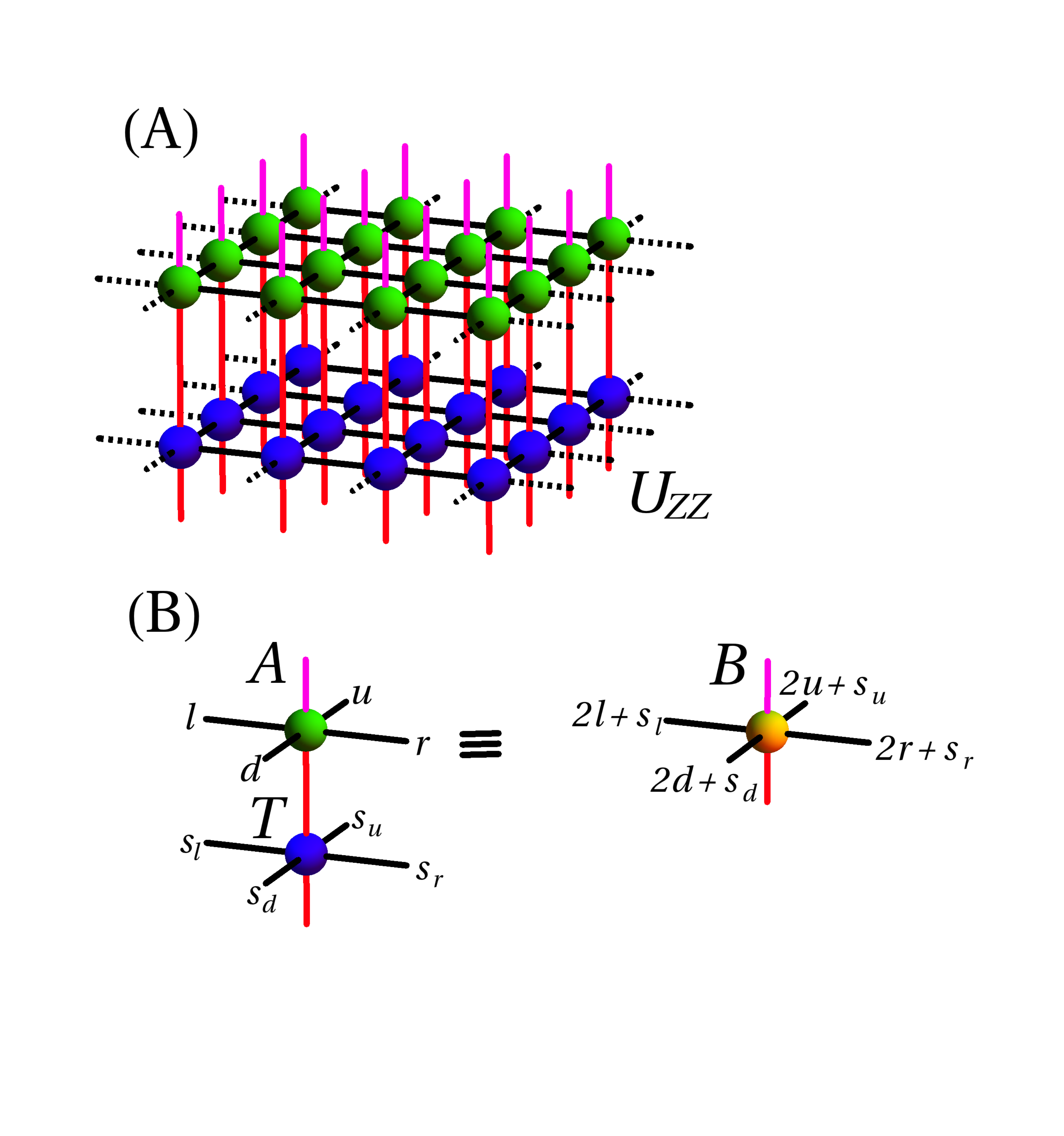}
\vspace{-1.9cm}
\caption{ 
In A,
the action of $U_{ZZ}(\beta)$ on the PEPS state in Fig. \ref{FigPeps}B.
Here $U_{ZZ}$ is represented by a tensor product operator being a contraction of elementary Trotter tensors $T$.
In B,
each PEPS tensor $A$ can be fused with its Trotter tensor into a composite tensor $B$.
Consequently,
the application $U_{ZZ}$ to the original PEPS made of tensors $A$ results 
in a new PEPS made of tensors $B$ with a bond dimension $2D$.
Its bond indices have to be renormalized back to the original dimension $D$. 
}
\label{FigB}
\end{figure}

The operator $U_{ZZ}(d\beta)$ is a product over nearest-neighbor bonds
\bea 
&&
U_{ZZ}(d\beta)=
\prod_{\langle m,m'\rangle}e^{\frac{d\beta}{2}Z_mZ_{m'}}=\nonumber\\
&&
\prod_{\langle m,m'\rangle}
\cosh\frac{d\beta}{2}
\sum_{s_{\langle m,m' \rangle}=0,1}
\Big({\cal O}_m {\cal O}_{m'}\Big)^{s_{\langle m,m' \rangle}}
\label{svd}
\eea
Here at each site $m$ we introduced ${\cal O}_m=Z_m\tanh^{1/2}\frac{d\beta}{2}$ and 
at each bond $\langle m,m' \rangle$ a bond index $s_{\langle m,m' \rangle}$.
With the help of the bond indices, 
$U_{ZZ}(d\beta)$ can be rearranged as a tensor product operator,
see Fig. \ref{FigB},
being a contraction of Trotter tensors,
\be 
T^{ij}_{s_u,s_r,s_d,s_l}(d\beta)=\cosh^2\frac{d\beta}{2}~\left({\cal O}^s\right)^{ij},
\label{T}
\ee 
through their bond indices $s_u,s_r,s_d,s_l\in\left\{0,1\right\}$. 
Here $s=s_u+s_r+s_d+s_l$ is a sum of bond indices coming out from a given site. 
The action of the tensor product operator $U_{ZZ}(d\beta)$ maps $A$ to a new tensor
\bea
&& 
B^{ia}_{2u+s_u,2r+s_r,2d+s_d,2l+s_l} = 
\sum_{j=0,1} 
T^{ij}_{s_u,s_r,s_d,s_l}~A^{ja}_{urdl}
\label{B}
\eea
see Fig. \ref{FigB}B.
This is an exact map, but $B$ has the bond dimension $2D$ instead of the original $D$. 

%%%%%%%%%%%%%%%%%%%%%%%%%%%%%%%%%%%%%%%%%%%%%%%%%%%%%%%%%%%%%%%%%%%%%%%%%%%%
%%%%%%%%%%%%%%%%%%%%%%%%%%%%%%%%%%%%%%%%%%%%%%%%%%%%%%%%%%%%%%%%%%%%%%%%%%%%

%%%%%%%%%%%%%%%%%%%%%%%%%%%%%%%%%%%%%%%%%%%%%%%%%%%%%%%%%%%%%%%%%%%%%%%%%% 
\section{ Tensor renormalization\label{SecW} }  
%%%%%%%%%%%%%%%%%%%%%%%%%%%%%%%%%%%%%%%%%%%%%%%%%%%%%%%%%%%%%%%%%%%%%%%%%% 

\begin{figure}[h!]
\vspace{-0.3cm}
\includegraphics[width=1.0\columnwidth,clip=true]{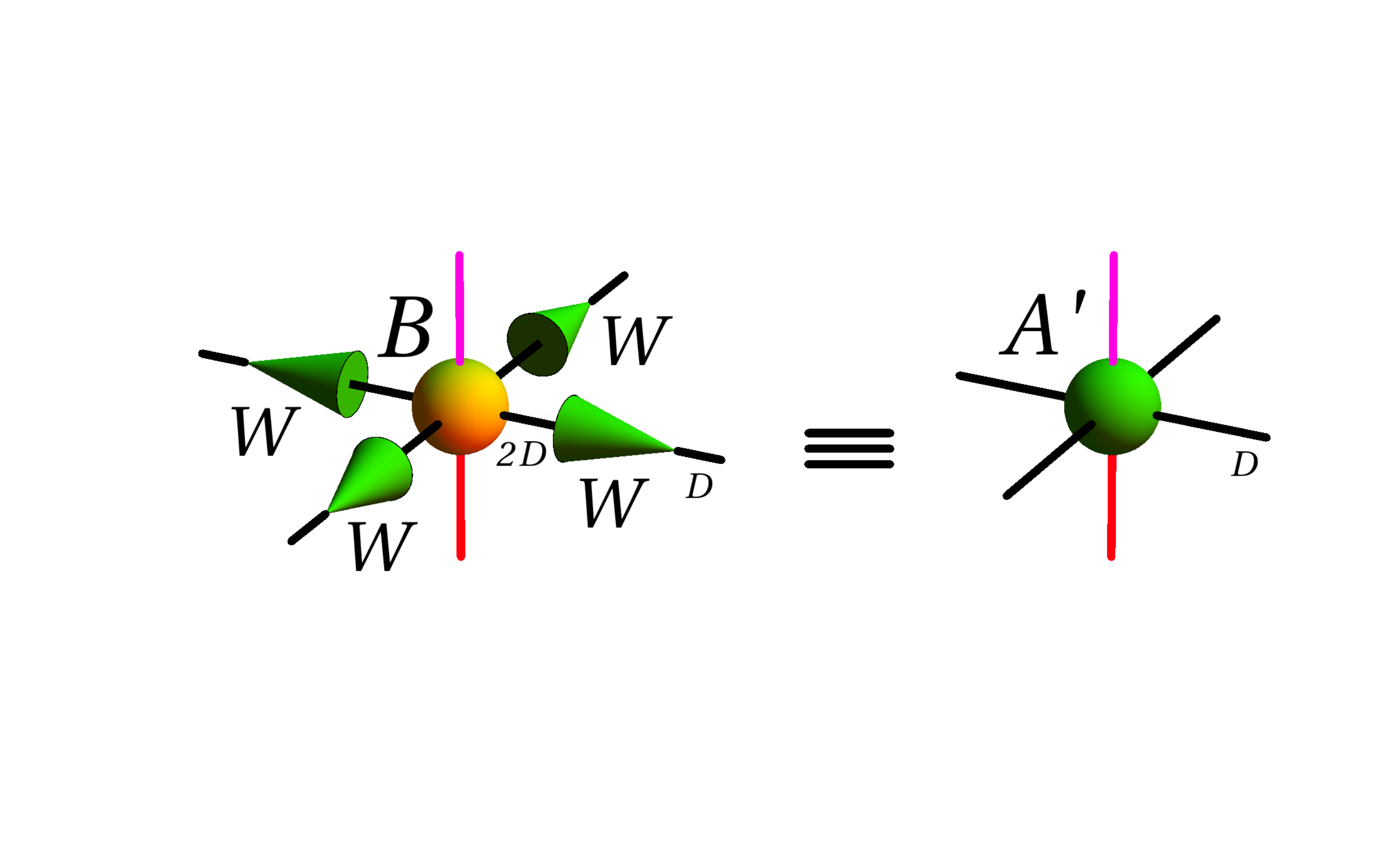}
\vspace{-1.5cm}
\caption{ 
The action of the isometry $W$ on the exact new tensor $B$,
compare Eq. (\ref{W}). 
The isometry truncates
the bond dimension from $2D$ back to the original $D$.
}
\label{FigAprime}
\end{figure}

The bond dimension has to be truncated back to $D$ in a way least distortive to the exact new PEPS $|\psi_B\rangle$. 
This renormalization can be done with an isometry $W$ that maps from $2D$ back to $D$ dimensions:
\be 
\sum_{u',r',d',l'=0}^{2D-1}
W_u^{u'}~
W_r^{r'}~
W_d^{d'}~
W_l^{l'}~
B^{ia}_{u'r'd'l'}~=~
A'^{ia}_{urdl}~,
\label{W}
\ee 
see Fig. \ref{FigAprime}. 
Here $A'$ is a candidate for a new tensor after the infinitesimal time step with bond indices $u,r,d,l=0,...,D-1$. 
The isometry is a variational parameter and 
the renormalized new PEPS $|\psi_{A'}\rangle$ is a function of this parameter.
$W$ should maximize a fidelity 
\be 
F=\langle\psi_{A'}|\psi_B\rangle
\label{F}
\ee 
between the exact $|\psi_B\rangle$ and the renormalized $|\psi_{A'}\rangle$.
From the point of view of numerical efficiency
the fidelity has a disadvantage that 
it involves tensor $B$ with the doubled bond dimension. 
In the following,
we optimize $A'$ avoiding this overhead,
but without compromising the accuracy of the second-order 
Suzuki-Trotter decomposition.

%%%%%%%%%%%%%%%%%%%%%%%%%%%%%%%%%%%%%%%%%%%%%%%%%%%%%%%%%%%%%%%%%%%%%%%%%%%%
%%%%%%%%%%%%%%%%%%%%%%%%%%%%%%%%%%%%%%%%%%%%%%%%%%%%%%%%%%%%%%%%%%%%%%%%%%%%

%%%%%%%%%%%%%%%%%%%%%%%%%%%%%%%%%%%%%%%%%%%%%%%%%%%%%%%%%%%%%%%%%%%%%%%%%% 
\section{ Self-consistent optimization \label{SecSelf}}  
%%%%%%%%%%%%%%%%%%%%%%%%%%%%%%%%%%%%%%%%%%%%%%%%%%%%%%%%%%%%%%%%%%%%%%%%%% 

The maximization of the fidelity (\ref{F}) with respect to $W$ can be done iteratively in 
a self-consistent way. We choose one bond in the bra $\langle\psi_{A'}|$ and optimize the 
isometry on this bond keeping the isometries on all other bonds fixed. Once the chosen 
isometry is optimized, it is used to replace isometries on all other bonds as well. This 
optimization is repeated until convergence. Finally, the PEPS tensor $A'$ in Eq. (\ref{W}) 
with the converged $W$ is accepted as the new tensor after the infinitesimal time step.   

In order to get rid of the numerical overhead introduced by tensor $B$,
we note that when $W$ is already close enough to the optimal one, 
then $F$ will not be distorted much 
when we replace $|\psi_B\rangle$ with $|\psi_{A'}\rangle$.
In fact, when $D$ is large enough and $W$ is optimal, 
then it will not be distorted at all. 
After this replacement we arrive at a new figure of merit:
\be 
\langle \psi_{A'} | \psi_{A'} \rangle.
\label{tildeF}
\ee
This is the norm of the PEPS with tensor $A'$,
but its maximization proceeds the same steps as the maximization of the fidelity (\ref{F}). 
We choose one bond in the bra $\langle \psi_{A'}|$ and optimize
the isometry on this bond keeping isometries on all other bonds and the isometries
in the ket $|\psi_{A'}\rangle$ fixed. Once the chosen isometry is optimized, it is used
to replace all other isometries as well, both in the bra and the ket. This optimization
is repeated until convergence. Tensor $A'$ with the converged $W$ is accepted as the
new PEPS tensor after the infinitesimal time step.

%%%%%%%%%%%%%%%%%%%%%%%%%%%%%%%%%%%%%%%%%%%%%%%%%%%%%%%%%%%%%%%%%%%%%%%%%%%%
\begin{figure}[h!]
\vspace{-0.3cm}
\includegraphics[width=1.0\columnwidth,clip=true]{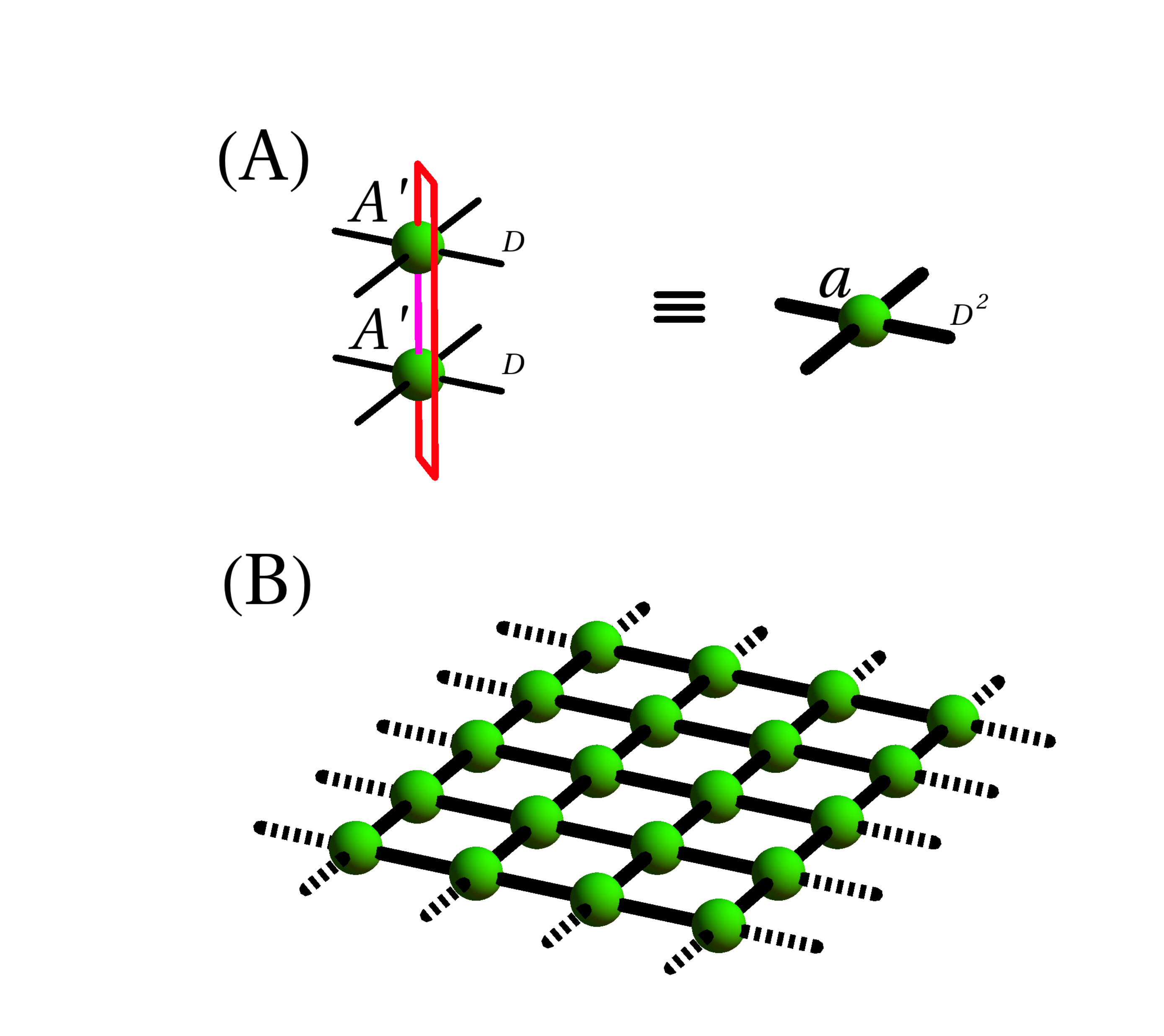}
\vspace{-0.8cm}
\caption{ 
In A,
tensor $A'$ is contracted with its conjugate through their spin and ancilla indices.
The contraction is a transfer tensor $a$.
In B,
a contraction of the transfer tensors is the norm $\langle\psi_{A'}|\psi_{A'}\rangle$ of 
the PEPS with tensor $A'$.
This infinite contraction can be done approximately with 
the help of the corner matrix renormalization \cite{CMR},
see Appendix \ref{CMR}.
}
\label{Figa}
\end{figure}
%%%%%%%%%%%%%%%%%%%%%%%%%%%%%%%%%%%%%%%%%%%%%%%%%%%%%%%%%%%%%%%%%%%%%%%%%%%%

\begin{figure}[h!]
\vspace{-0.3cm}
\includegraphics[width=1.0\columnwidth,clip=true]{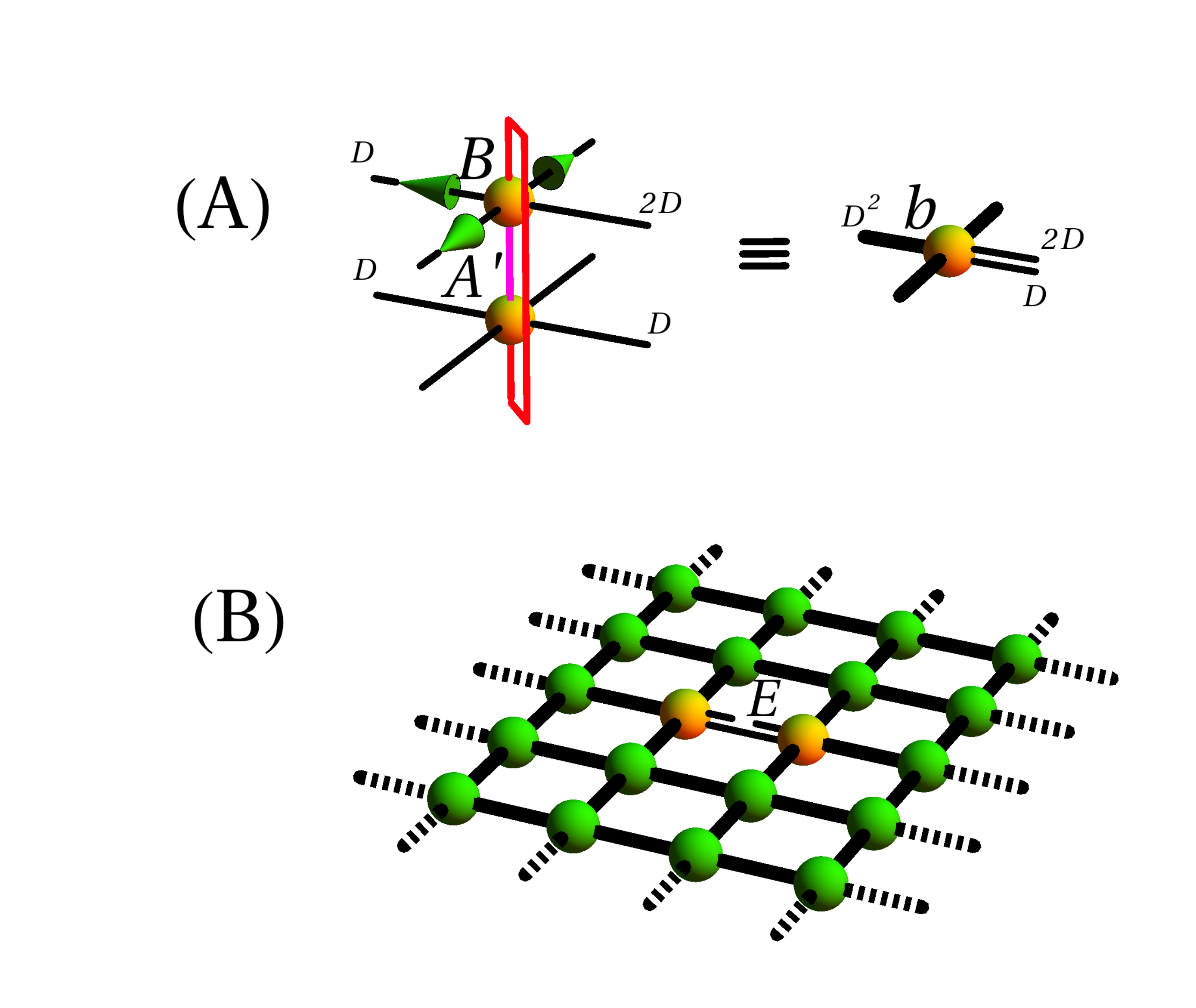}
\vspace{-0.8cm}
\caption{ 
In A,
a transfer tensor $b$ on a site at the chosen bond. In contrast to tensor $a$
in Fig. \ref{Figa}A, one of its bond indices along the chosen bond (the one 
within the bra) is left unrenormalized by the isometry. 
In B,
the norm in Fig. \ref{Figa}B but with the chosen bond in the bra left 
uncontracted and its bond indices unrenormalized. This is a bond environment
matrix $E$. 
This infinite tensor network is contracted by a method described 
in Appendix \ref{CMR}.
}
\label{FigE}
\end{figure}

Having outlined the algorithm, we still need to specify how to optimize the isometry
on the chosen bond. 
To begin with, we note that a tensor network representing the norm (\ref{tildeF}) 
can be obtained as in Fig. \ref{Figa}. 
However, what we actually need is this norm, but with the chosen bond in the bra left 
uncontracted and its bond indices left unrenormalized by the isometry.
This object can be obtained as in Fig. \ref{FigE}. 
We call it a ``bond environment matrix'' $E^{kl}$.
Its indices are the unrenormalized and uncontracted bond indices along the chosen bond.
From $E^{kl}$ we can obtain the norm by overlapping it with the projector 
$P^{kl}=\sum_{b=0}^{D-1} W^k_b W^l_b$ made of the isometry $W$:
\be 
\langle \psi_{A'} | \psi_{A'} \rangle=
\sum_{k,l=0}^{2D-1}P^{kl}E^{kl}=
\sum_{b=0}^{D-1}  \sum_{k,l=0}^{2D-1} W^k_b W^l_b E^{kl}.
\label{fom}
\ee
Here the isometries renormalize the uncontracted bond indices $k,l$ 
and then the sum over $b$ contracts the renormalized ones. 

The norm (\ref{fom}) has to be maximized with respect to $W$ at fixed $E$.
To this end, we diagonalize the symmetric $E$:
\be 
E^{kl}=\sum_{b=0}^{2D-1} V^k_b \lambda_b V^l_b
\label{ediag}
\ee
with eigenvalues $\lambda_0\geq...\geq\lambda_{2D-1}$. 
The optimal isometry is made of the leading eigenvectors:
\be
W^k_b=V^k_b
\ee 
for $b=0,...,D-1$. It maximizes the figure of merit (\ref{fom}) as
$\langle \psi_{A'}|\psi_{A'} \rangle=\sum_{b=0}^{D-1}\lambda_b$.

Finally, we note that since the projector $P^{kl}$ is symmetric, in general 
only the symmetric part of $E^{kl}$ contributes to the norm (\ref{fom}) and, 
consequently, only the symmetric part has to be diagonalized in Eq. (\ref{ediag}).
This observation is used in Section \ref{SecFinite} below.

%%%%%%%%%%%%%%%%%%%%%%%%%%%%%%%%%%%%%%%%%%%%%%%%%%%%%%%%%%%%%%%%%%%%%%%%%%%%
%%%%%%%%%%%%%%%%%%%%%%%%%%%%%%%%%%%%%%%%%%%%%%%%%%%%%%%%%%%%%%%%%%%%%%%%%%%%

%%%%%%%%%%%%%%%%%%%%%%%%%%%%%%%%%%%%%%%%%%%%%%%%%%%%%%%%%%%%%%%%%%%%%%%%%% 
\section{ Benchmark results \label{SecBenchmark}}  
%%%%%%%%%%%%%%%%%%%%%%%%%%%%%%%%%%%%%%%%%%%%%%%%%%%%%%%%%%%%%%%%%%%%%%%%%% 

In order to evolve PEPS across the symmetry breaking phase transition
we add to the Hamiltonian (\ref{calH}) a tiny longitudinal bias 
\be 
{\cal H}_Z = - \delta \sum_s Z_s 
\ee 
to smooth out the transition at a finite $\beta_c(h)$. 
We present results for the transverse field $h=\frac23h_c$
that is large enough to introduce substantial quantum fluctuations and 
significantly increase $\beta_c$ above the Onsager's $\beta_0$.

\begin{figure}[h!]
%\vspace{-0.5cm}
\includegraphics[width=0.99\columnwidth,clip=true]{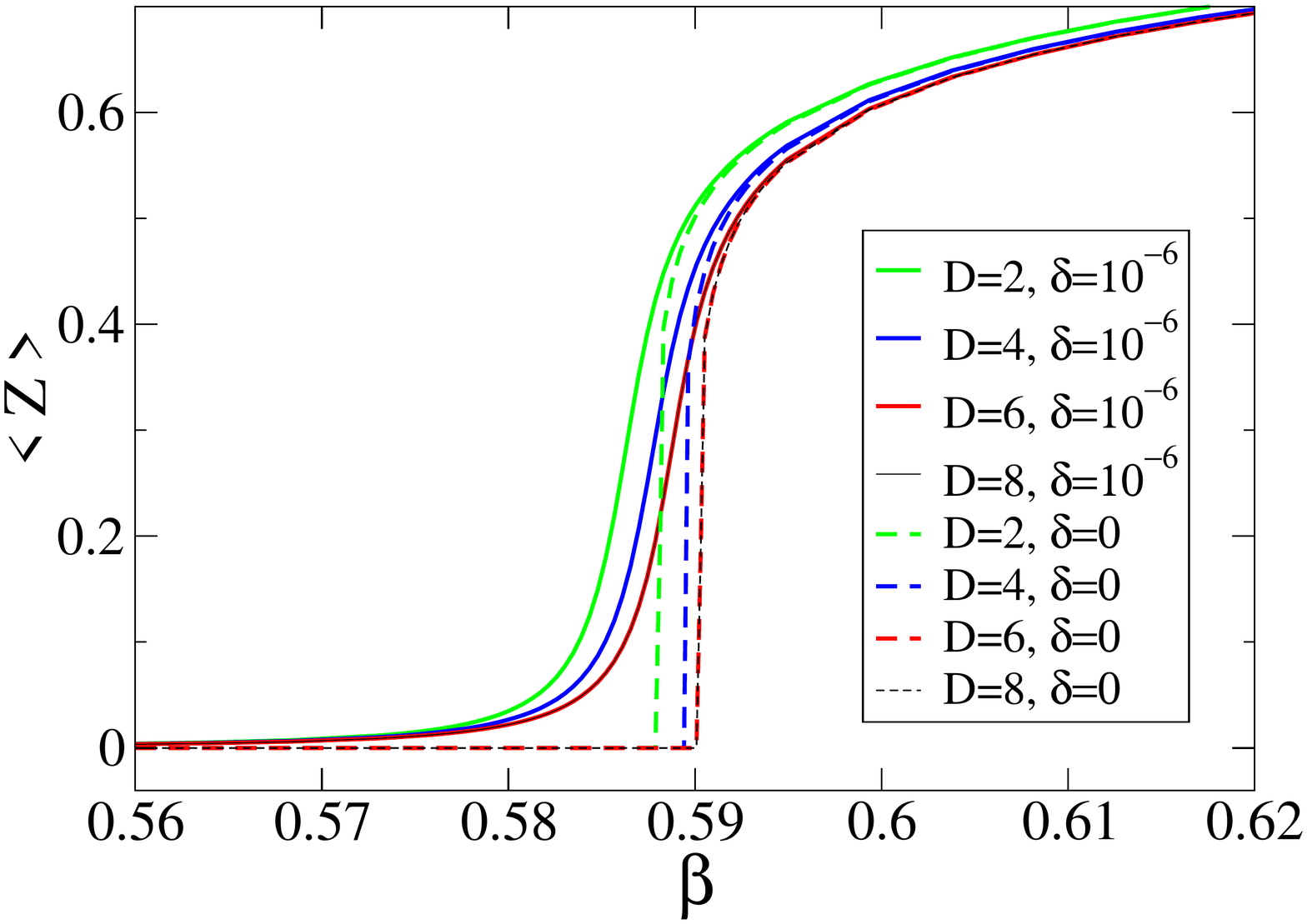}
\vspace{-0.2cm}
\caption{ 
The ferromagnetic magnetization $\langle Z \rangle$ as a function of $\beta$ for 
the transverse field $h=\frac23h_0$ and the longitudinal bias $\delta=10^{-6}$.
This magnetization and other local observables are converged for $D\geq6$
and $M\geq12$. Here we show results for $M=16$. 
For comparison, 
we also show results without the bias field, $\delta=0$. 
Here the time step near $\beta_c$ is $d\beta=10^{-3}\beta_0=0.000441$.
}
\label{FigZ}
\end{figure}

Figure \ref{FigZ} shows the order parameter $\langle Z \rangle$ as a function of $\beta$
for $\delta=10^{-6}$. The slope $d\langle Z\rangle/d\beta$ is the steepest at $\beta_c=0.589=1.33\beta_0$. 
This local observable requires $D\geq6$ and a relatively small environmental bond dimension, $M\geq12$, 
to converge. 

\begin{figure}[h!]
%\vspace{-0.5cm}
\includegraphics[width=1.0\columnwidth,clip=true]{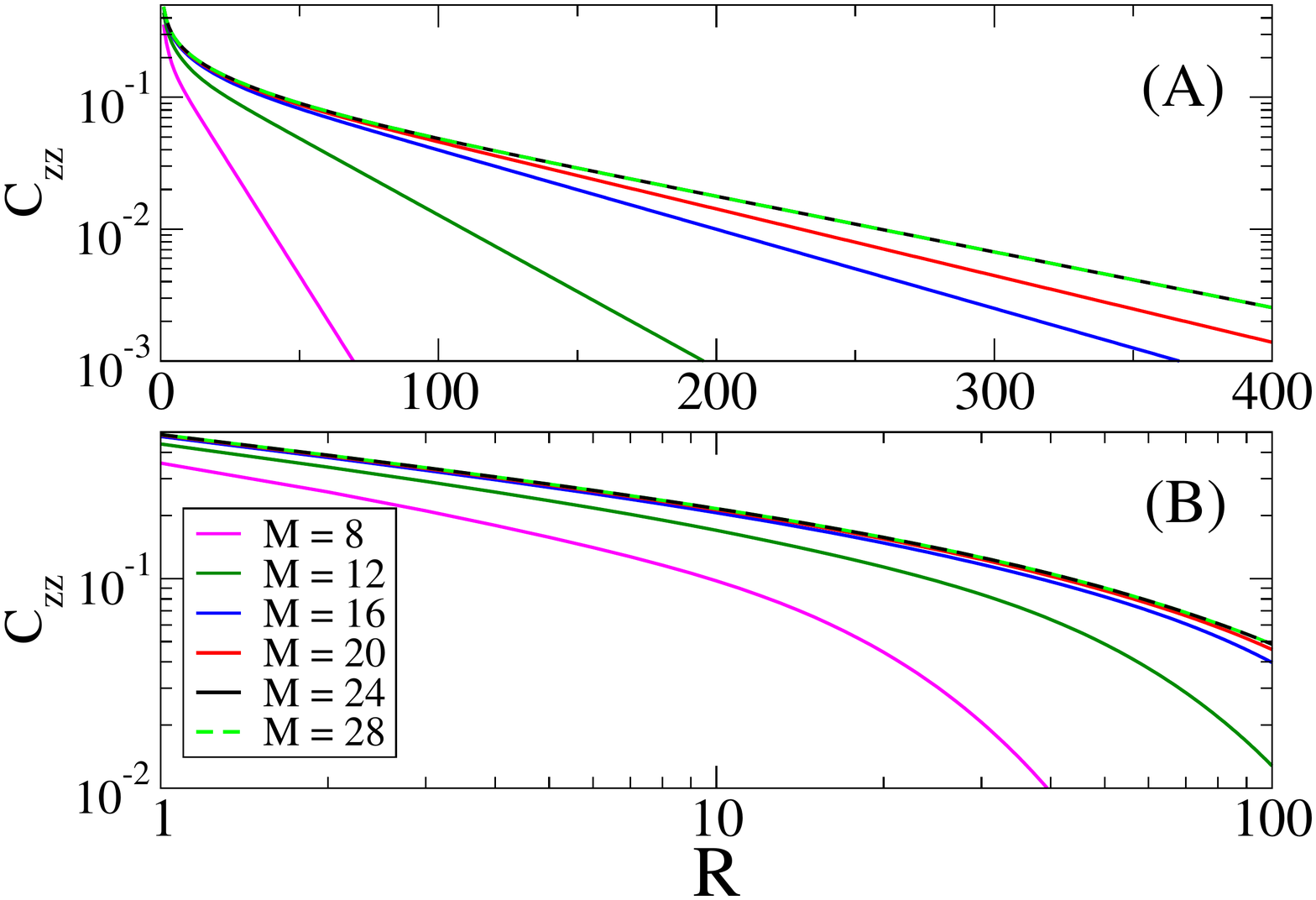}
\vspace{-0.2cm}
\caption{ 
The ferromagnetic correlator, 
$
C_{zz}(R)=
\langle Z_{(x+R,y)} Z_{(x,y)}\rangle-
\langle Z \rangle^2
$,
for the transverse field $h=\frac23h_c$, 
the longitudinal bias $h_Z=10^{-6}$,
at the critical $\beta_c=0.589$.
The correlator is converged for $M\geq24$ and $D\geq6$. 
Here we show $D=6$.
In A,
a logarithmic plot emphasizing exponential tails for large $R$:
$
C_{zz}(R) \sim e^{-R/\xi}.
$
The correlation length increases with $M$ until it converges at $\xi=224$
which is finite thanks to the finite bias.
In B,
a log-log plot emphasizing a power law decay for intermediate $R$, 
$
C_{zz}(R) \sim R^{-\eta}.
$  
The best fit is $\eta=0.28$.
}
\label{FigCzz}
\end{figure}

Long range correlations are more demanding on $M$,
as demonstrated by the ferromagnetic correlator in Fig. \ref{FigCzz}.
For $\delta=10^{-6}$ 
we need $M\geq20$ to converge a finite correlation length $\xi\approx224$.
It is finite thanks to the bias.
As shown in Appendix \ref{SecOnsager},
$\delta=0$ would make both $\xi$ and necessary $M$ diverge at the critical point 
making accurate evolution across $\beta_c$ impossible.

Indeed, in Fig. \ref{FigZ} we also show $\langle Z \rangle$ obtained with $\delta=0$. 
Here the lack of convergence in $M$ manifests itself by a discontinuous jump in 
the spontaneous magnetization. Nevertheless, deeper in the ferromagnetic phase, 
where the influence of the bias becomes negligible, the magnetization overlaps 
with the smooth curve obtained with the tiny bias and converged in $M$.

%%%%%%%%%%%%%%%%%%%%%%%%%%%%%%%%%%%%%%%%%%%%%%%%%%%%%%%%%%%%%%%%%%%%%%%%%%%%
%%%%%%%%%%%%%%%%%%%%%%%%%%%%%%%%%%%%%%%%%%%%%%%%%%%%%%%%%%%%%%%%%%%%%%%%%%%%

%%%%%%%%%%%%%%%%%%%%%%%%%%%%%%%%%%%%%%%%%%%%%%%%%%%%%%%%%%%%%%%%%%%%%%%%%%%%
%%%%%%%%%%%%%%%%%%%%%%%%%%%%%%%%%%%%%%%%%%%%%%%%%%%%%%%%%%%%%%%%%%%%%%%%%%%%

%%%%%%%%%%%%%%%%%%%%%%%%%%%%%%%%%%%%%%%%%%%%%%%%%%%%%%%%%%%%%%%%%%%%%%%%%% 
\section{ Finite lattice \label{SecFinite}}  
%%%%%%%%%%%%%%%%%%%%%%%%%%%%%%%%%%%%%%%%%%%%%%%%%%%%%%%%%%%%%%%%%%%%%%%%%% 

\begin{figure}[h!]
%\vspace{-0.5cm}
\includegraphics[width=0.8\columnwidth,clip=true]{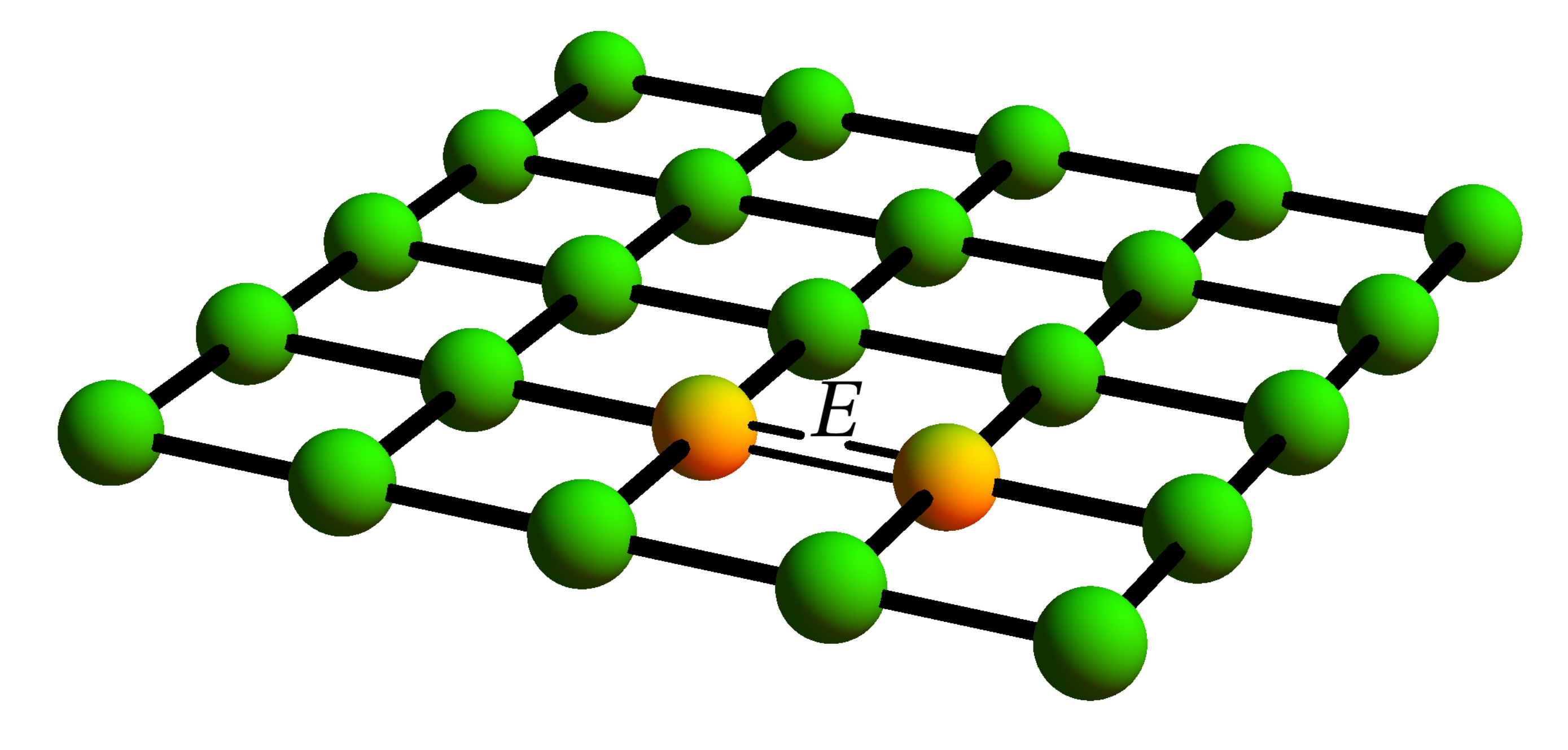}
\vspace{-0.2cm}
\caption{ 
One of the bond environments on a $5\times5$ lattice.
}
\label{FigEfinite}
\end{figure}

On a finite $N\times N$ lattice 
the self-consistent algorithm is basically the same as on the infinite one except that 
contraction of finite networks, 
like the one in Fig. \ref{FigEfinite}, 
can be done with matrix product states \cite{OPS} instead of the corner matrix renormalization in Appendix \ref{CMR}. 
A finite lattice is also less symmetric than the infinite one,
hence different bonds and their environments $E$ are not equivalent.
One has to sweep across the lattice optimizing each bond separately -
modulo residual symmetries -
and repeat the sweeps until convergence of all bonds.

%%%%%%%%%%%%%%%%%%%%%%%%%%%%%%%%%%%%%%%%%%%%%%%%%%%%%%%%%%%%%%%%%%%%%%%%%%%%
%%%%%%%%%%%%%%%%%%%%%%%%%%%%%%%%%%%%%%%%%%%%%%%%%%%%%%%%%%%%%%%%%%%%%%%%%%%%

\begin{figure}[h!]
%\vspace{-0.5cm}
\includegraphics[width=0.99\columnwidth,clip=true]{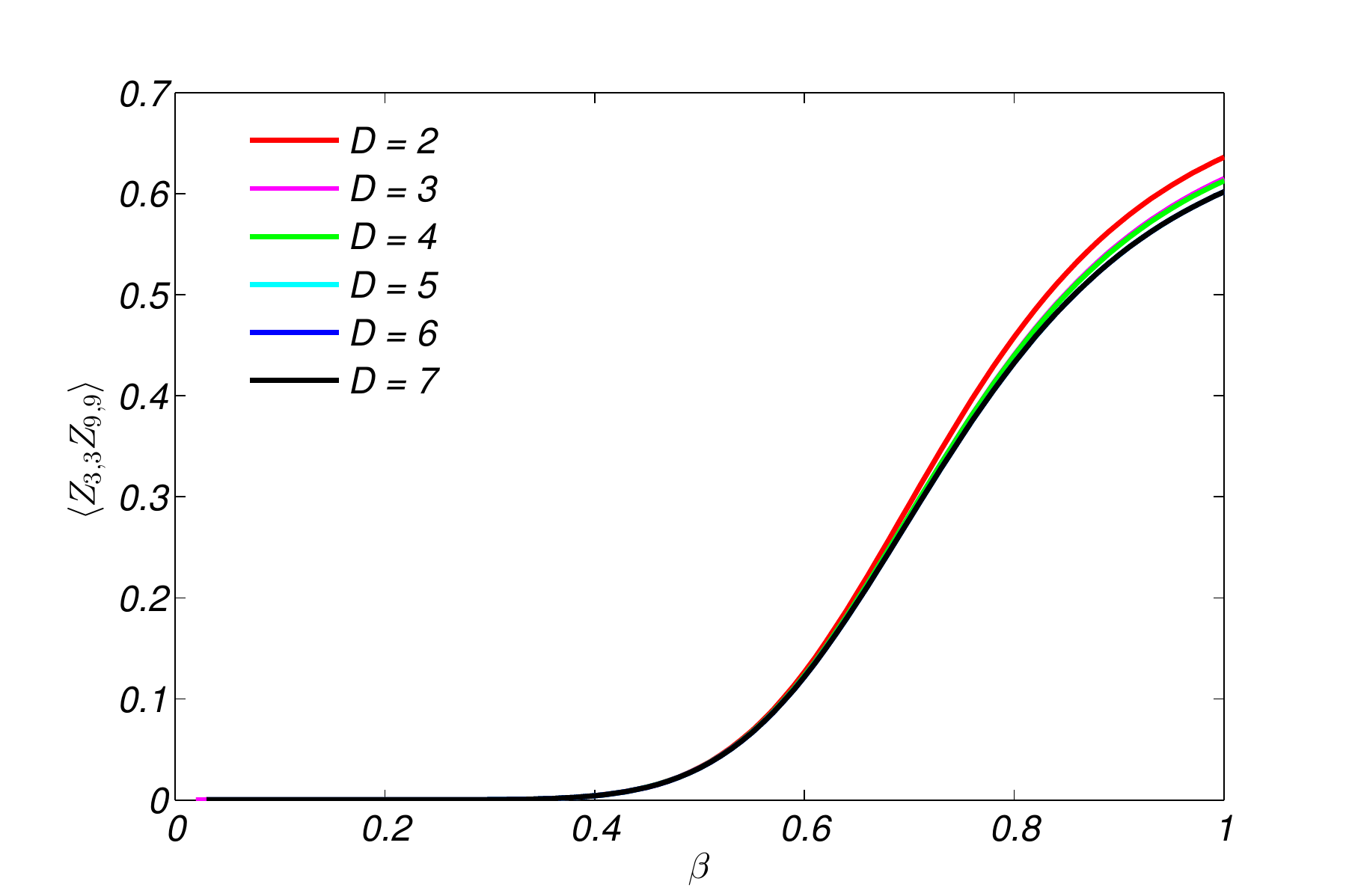}
\vspace{-0.2cm}
\caption{ 
Ferromagnetic correlator $<Z_{3,3}Z_{9,9}>$ between sites $(3,3)$ and $(9,9)$
on a diagonal of an $11\times11$ lattice as a function of $\beta$. 
Here the transverse field $g=\frac23g_c$,
the longitudinal bias $\delta=0$,
and a bond dimension in the matrix product state is $M_{\rm MPS}=16$.
The correlator is converged for bond dimension $D\geq5$ and $M_{\rm MPS}  \geq16$. 
}
\label{FigResFinite}
\end{figure}

Figure \ref{FigResFinite} shows a ferromagnetic correlator along a diagonal of an $11\times11$ lattice. 
The finite lattice size itself is enough to smooth the transition, hence there is no need for 
the longitudinal bias here and we set $\delta=0$.  

%%%%%%%%%%%%%%%%%%%%%%%%%%%%%%%%%%%%%%%%%%%%%%%%%%%%%%%%%%%%%%%%%%%%%%%%%%%%
%%%%%%%%%%%%%%%%%%%%%%%%%%%%%%%%%%%%%%%%%%%%%%%%%%%%%%%%%%%%%%%%%%%%%%%%%%%%

%%%%%%%%%%%%%%%%%%%%%%%%%%%%%%%%%%%%%%%%%%%%%%%%%%%%%%%%%%%%%%%%%%%%%%%%%% 
\section{ Conclusion \label{SecConclusion}} 
%%%%%%%%%%%%%%%%%%%%%%%%%%%%%%%%%%%%%%%%%%%%%%%%%%%%%%%%%%%%%%%%%%%%%%%%%% 

We presented an iterative self-consistent renormalization of the PEPS bond dimension after a Suzuki-Trotter gate.
The procedure takes into account full environment made of self-consistently renormalized tensors to preserve the 
accuracy of the  Suzuki-Trotter decomposition. The algorithm is efficient: it takes one day in MATLAB on a laptop 
computer to obtain the data in this paper. Its formal cost - dominated by the corner matrix renormalization (CMR) - 
scales like $M^3D^6$, but in more efficient versions of CMR \cite{CMR} it can be cut down to ${\cal O}(M^3D^2)$.
In conclusion, the protocol brings us closer to a numerically exact imaginary time evolution that can be used 
to obtain thermal states of strongly correlated systems. 

%%%%%%%%%%%%%%%%%%%%%%%%%%%%%%%%%%%%%%%%%%%%%%%%%%%%%%%%%%%%%%%%%%%%%%%%%%%%% 
\acknowledgements
%%%%%%%%%%%%%%%%%%%%%%%%%%%%%%%%%%%%%%%%%%%%%%%%%%%%%%%%%%%%%%%%%%%%%%%%%%%%%
This work was supported by the Polish National Science Center (NCN) under Project DEC-2013/09/B/ST3/01603.
%%%%%%%%%%%%%%%%%%%%%%%%%%%%%%%%%%%%%%%%%%%%%%%%%%%%%%%%%%%%%%%%%%%%%%%%%%%%%

%%%%%%%%%%%%%%%%%%%%%%%%%%%%%%%%%%%%%%%%%%%%%%%%%%%%%%%%%%%%%%%%%%%%%%%%%%%%%%%%%%%%%%%%%%%%%%%%%%%%%%%%%%%%%%%%%%

%%%%%%%%%%%%%%%%%%%%%%%%%%%%%%%%%%%%%%%%%%%%%%%%%%%%%%%%%%%%%%%%%%%%%%%%%%%%%%%%%%%%%%%%%
\appendix
%%%%%%%%%%%%%%%%%%%%%%%%%%%%%%%%%%%%%%%%%%%%%%%%%%%%%%%%%%%%%%%%%%%%%%%%%%%%%%%%%%%%%%%%%
\vspace*{0.5cm}

%%%%%%%%%%%%%%%%%%%%%%%%%%%%%%%%%%%%%%%%%%%%%%%%%%%%%%%%%%%%%%%%%%%%%%%%%%%%%%%%%%%%%%%%%
\section{Corner matrix renormalization\label{CMR}}
%%%%%%%%%%%%%%%%%%%%%%%%%%%%%%%%%%%%%%%%%%%%%%%%%%%%%%%%%%%%%%%%%%%%%%%%%%%%%%%%%%%%%%%%%

\begin{figure}[h!]
%\vspace{-0.5cm}
\includegraphics[width=0.99\columnwidth,clip=true]{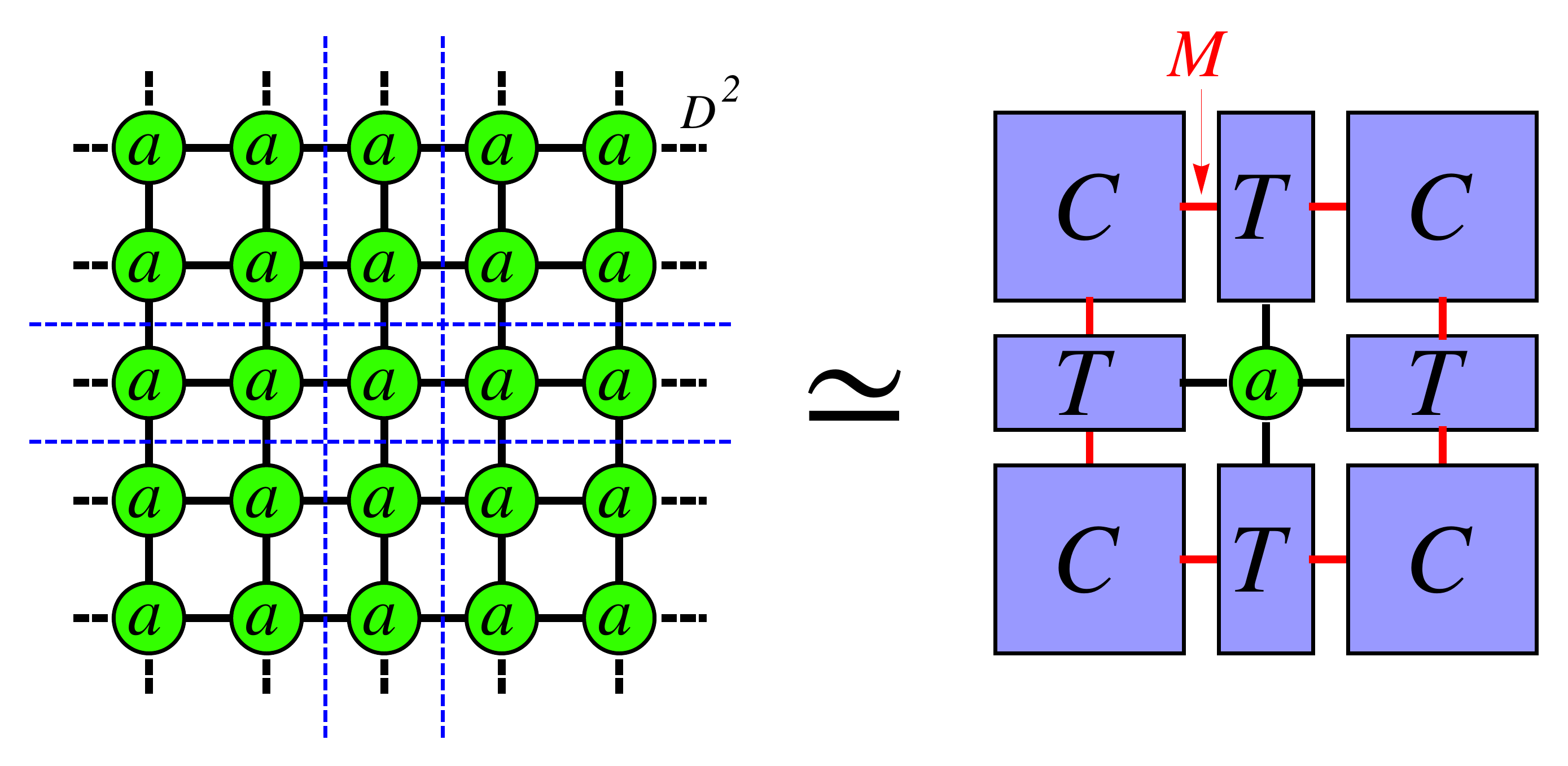}
\vspace{-0.2cm}
\caption{ 
On the left,
the norm 
$
\langle\psi_A|\psi_A\rangle
$
in Fig. \ref{Figa}B.
This infinite contraction cannot be done exactly,
hence it is approximated by the finite network on the right.
Corner matrices $C$ and top tensors $T$ effectively represent corresponding infinite sectors of 
the network on the left separated by the dashed blue lines. 
Their (red) environmental bonds have dimension $M$. 
The environmental tensors $C$ and $T$ should be such that, 
to the transfer tensor $a$ in the center, 
its environment on the right appears the same as its exact environment on the left as much as possible.
They are obtained by iterating the corner matrix renormalization in Fig. \ref{FigRenC} until convergence.
}
\label{FigCV}
\end{figure}

An infinite tensor network, 
like the one on the left of Fig. \ref{FigCV}, 
cannot be contracted exactly.
Fortunately,
what we need in general is not this number,
but an environment for a few tensors of interest.
For instance, 
in Fig. \ref{FigCV} we want an environment for the transfer tensor $a$ in the center.
The environment is a tensor that remains after removing the central tensor from the infinite network.
From the point of view of the central tensor,
its environment can be substituted with an effective environment, 
made of finite corner matrices $C$ and top tensors $T$,
that appears to the central tensor the same as the exact environment as much as possible.
The environmental tensors are contracted with each other by indices of dimension $M$.   
Increasing $M$ makes the effective environment more accurate,
and for a finite correlation length the environment is expected to converge at a finite $M$. 
 
For the sake of simplicity,
here in the Ising model, 
we assume that tensors $A,B,a$ are isotropic in their four bond indices.
Consequently,
$C$ is symmetric and top is symmetric in its environmental indices.

\begin{figure}[h!]
%\vspace{-0.5cm}
\includegraphics[width=0.8\columnwidth,clip=true]{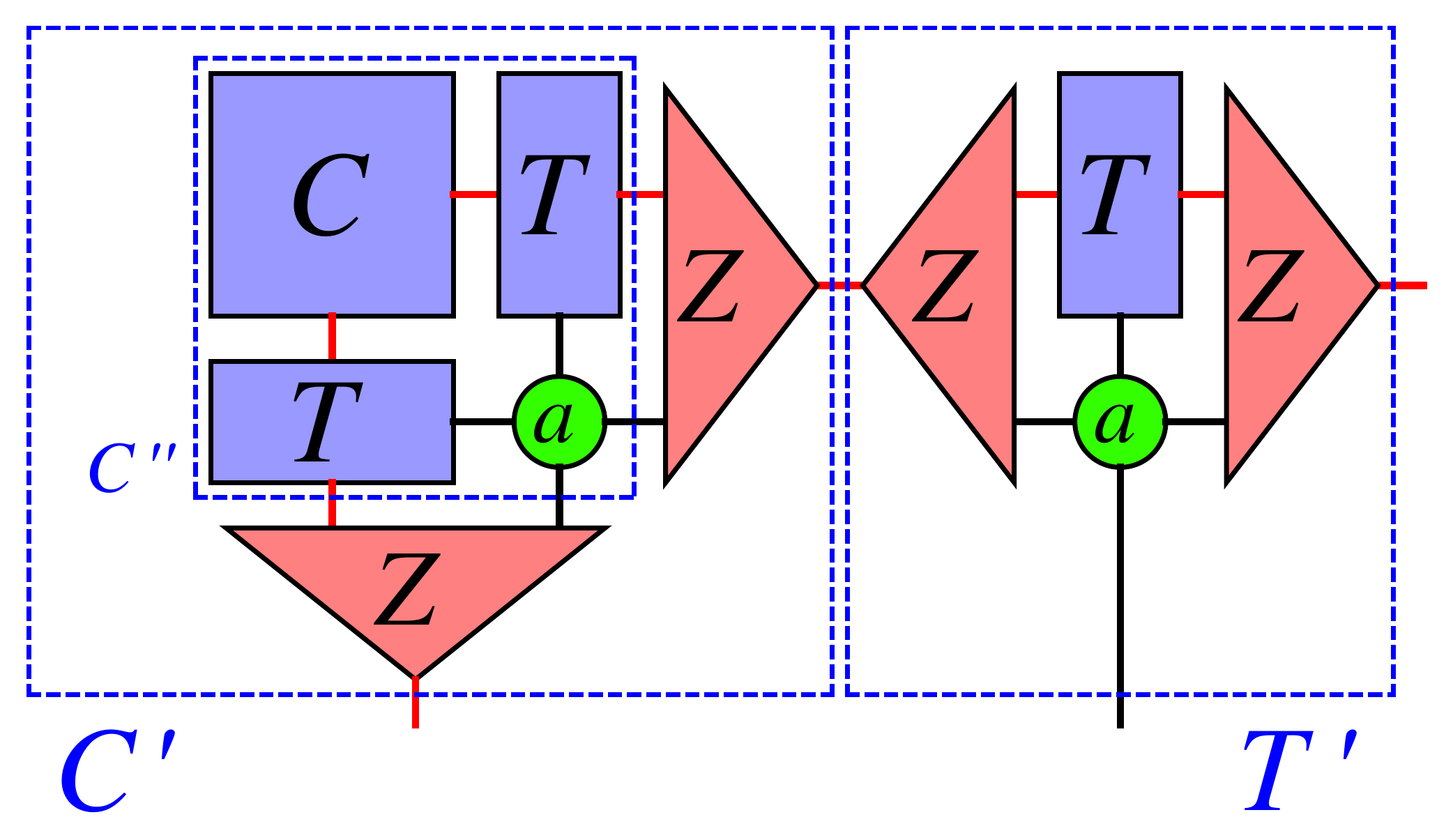}
\vspace{-0.2cm}
\caption{ 
The corner $C$ and top $T$ are obtained by repeating a renormalization procedure until convergence. 
The procedure has four steps. 
In the first step,
the tensors $C,T,a$ are contracted to an enlarged corner $C''$. 
In the second step,
the symmetric $MD^2\times MD^2$ matrix $C''$ is diagonalized and
its $M$ eigenvectors with the largest eigenvalues define an isometry $Z$. 
The diagonalization that scales like $M^3D^6$ is the leading cost
of this variant of corner matrix renormalization. 
In the third step,
$Z$ is used to renormalize/truncate the indices of $C''$ back to the original dimension $M$ 
giving a new (diagonal) corner $C'$.
In the fourth step, 
the same $Z$ renormalizes the contraction of $T$ with $a$ to a new $T'$. 
The four-step procedure is repeated until convergence of the $M$ leading eigenvalues of $C''$.
}
\label{FigRenC}
\end{figure}

Finite tensors $C$ and $T$ represent infinite sectors of the network on the left of Fig. \ref{FigCV}.
The tensors are converged by iterating the corner matrix renormalization in Fig. \ref{FigRenC}.
In every renormalization step,
the corner matrix is enlarged with one tensor $a$ and two $T$'s.
This operation represents 
the top-left corner sector in Fig. \ref{FigCV} absorbing one more layer of tensors $a$. 
Once the environment is converged,
it can be used to calculate efficiently either observables or the bond environment $E$,
see Fig. \ref{FigCTE}.
Initialization of the iterative corner renormalization is discussed in Appendix \ref{SecAcc}.

%%%%%%%%%%%%%%%%%%%%%%%%%%%%%%%%%%%%%%%%%%%%%%%%%%%%%%%%%%%%%%%%%%%%%%%%%%%%
%%%%%%%%%%%%%%%%%%%%%%%%%%%%%%%%%%%%%%%%%%%%%%%%%%%%%%%%%%%%%%%%%%%%%%%%%%%%

%%%%%%%%%%%%%%%%%%%%%%%%%%%%%%%%%%%%%%%%%%%%%%%%%%%%%%%%%%%%%%%%%%%%%%%%%%%%
%%%%%%%%%%%%%%%%%%%%%%%%%%%%%%%%%%%%%%%%%%%%%%%%%%%%%%%%%%%%%%%%%%%%%%%%%%%%

%%%%%%%%%%%%%%%%%%%%%%%%%%%%%%%%%%%%%%%%%%%%%%%%%%%%%%%%%%%%%%%%%%%%%%%%%%%%
\begin{figure}[h!]
%\vspace{-0.5cm}
\includegraphics[width=0.4\columnwidth,clip=true]{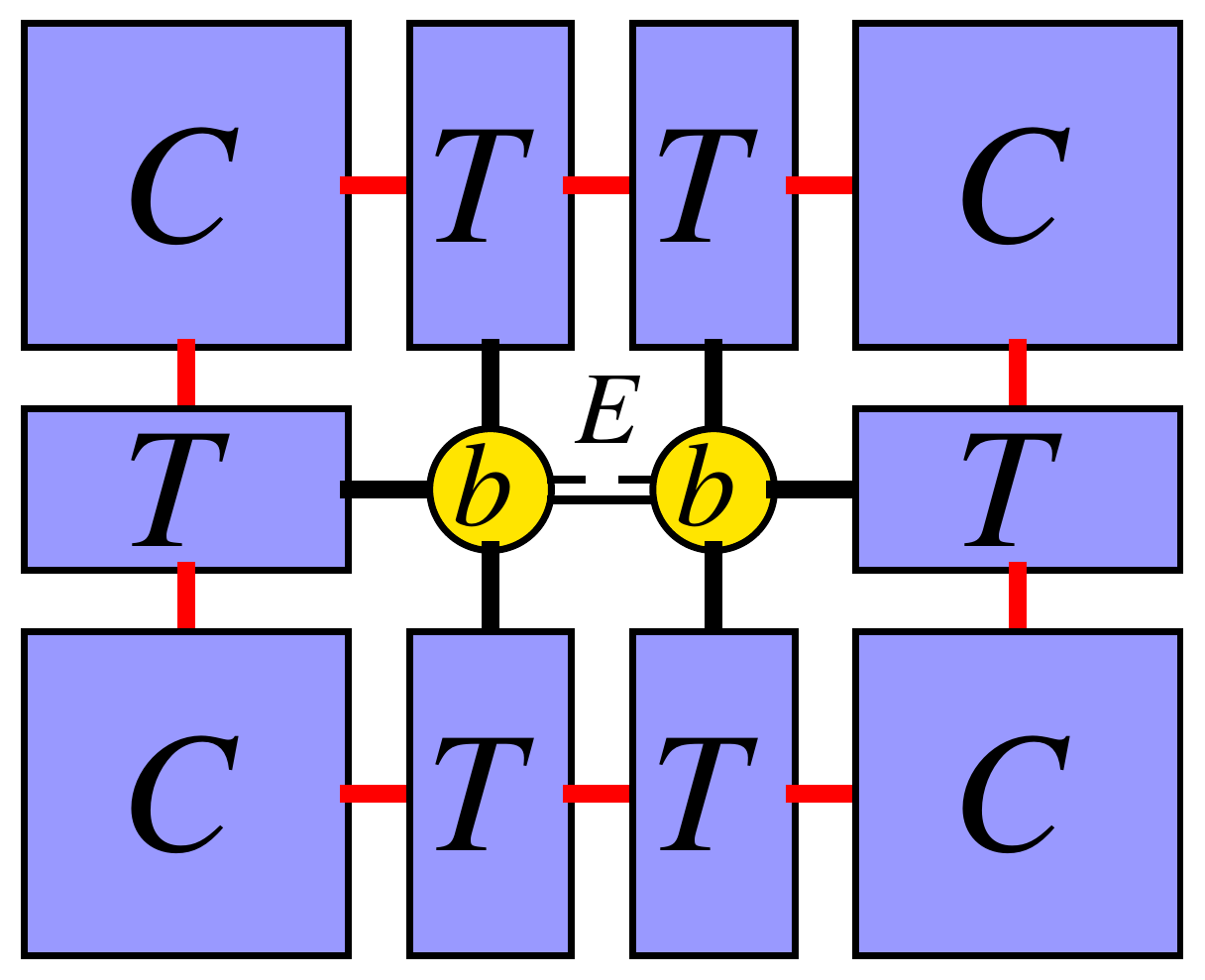}
\vspace{-0.2cm}
\caption{ 
A finite network approximating the exact bond environment $E$ in Fig. \ref{FigE}C.
For a finite correlation length, 
it is expected to become exact at a finite $M$.
}
\label{FigCTE}
\end{figure}
%%%%%%%%%%%%%%%%%%%%%%%%%%%%%%%%%%%%%%%%%%%%%%%%%%%%%%%%%%%%%%%%%%%%%%%%%%%%

%%%%%%%%%%%%%%%%%%%%%%%%%%%%%%%%%%%%%%%%%%%%%%%%%%%%%%%%%%%%%%%%%%%%%%%%%%%%%%%%%%%%%%%%%
\section{Gauge accelerator for corner matrix renormalization\label{SecAcc}}
%%%%%%%%%%%%%%%%%%%%%%%%%%%%%%%%%%%%%%%%%%%%%%%%%%%%%%%%%%%%%%%%%%%%%%%%%%%%%%%%%%%%%%%%%

%%%%%%%%%%%%%%%%%%%%%%%%%%%%%%%%%%%%%%%%%%%%%%%%%%%%%%%%%%%%%%%%%%%%%%%%%%%%
\begin{figure}[h!]
\vspace{+0.5cm}
\includegraphics[width=0.4\columnwidth,clip=true]{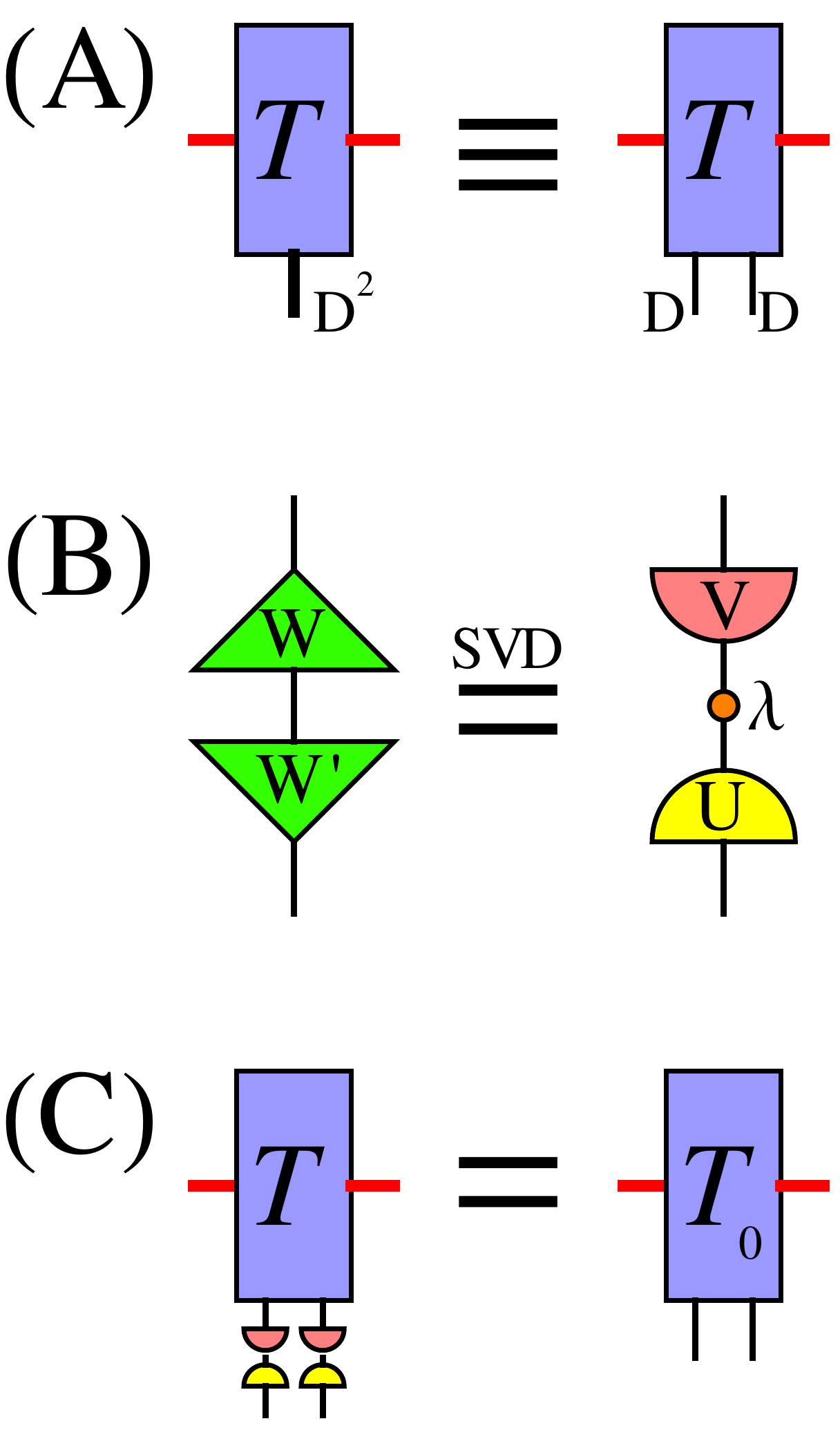}
\vspace{-0.2cm}
\caption{ 
In A,
the bottom index of tensor $T$ effectively comes out from tensor $a$,
see Figs. \ref{FigCV} and \ref{FigRenC}.
It is a fusion of two bond indices,
each of dimension $D$,
coming out from two tensors $A$,
see Fig. \ref{Figa}A.
Each bond index of tensor $A$ is actually a bond index of tensor $B$ renormalized by isometry $W$,
see Fig. \ref{FigAprime}.
In B,
an overlap between old isometry $W$ and new $W'$ is a $D\times D$ matrix. 
It can be subject to a singular value decomposition,
$W'W^T=U\lambda V^T$,
with singular values $\lambda$.
In C,
an orthogonal matrix $UV^T$ is applied to each bottom index of the old tensor $T$.
It is a gauge transformation that provides a better starting point $T_0$ for
the iterative procedure in Fig. \ref{FigRenC} converging the environmental tensors.
}
\label{FigAcc}
\end{figure}
%%%%%%%%%%%%%%%%%%%%%%%%%%%%%%%%%%%%%%%%%%%%%%%%%%%%%%%%%%%%%%%%%%%%%%%%%%%%

The iterative procedure in Fig. \ref{FigRenC} converging the environmental tensors is 
the most time-consuming part of the algorithm. 
It is accelerated by using in the environment the renormalized tensor $A$ instead of the full tensor $B$, 
but at the price of repeating the self-consistent iterative update of $A$ (or equivalently $W$). 
Before every update of $A$,
the environment has to be converged with current $A$.
Once $A$ is updated,
the environment has to be converged again with the updated $A$, 
and so forth until convergence of $A$.
Fortunately,
the convergence of the environment can be accelerated with a good choice of initial tensors $C_0$ and $T_0$.
When $A$ is already close to convergence and it is changed little by its update,
then the previous tensors $C$ and $T$, 
converged before the latest update of $A$,
can be reused as the initial tensors.
Moreover,
they can be improved even further at negligible cost.

The update changes tensor $A$ by updating the isometry from an old $W$ to a new $W'$.
As explained in Fig. \ref{FigAcc}A,
the bottom index of tensor $T$ is actually a fusion of two bond indices of two tensors $A$,
and these bond indices are in fact bond indices of tensors $B$ renormalized by the old isometry $W$.
We would like to replace this old $W$ with the new $W'$ but,
since $W$ is not invertible,
this cannot be done exactly.  
However,
as explained in Figs. \ref{FigAcc}B and C,
we can apply an orthogonal (gauge) transformation to the bottom index of $T$
that is as close to the exact replacement $W\to W'$ as possible.
The transformed $T_0$ is a better starting point than
the $T$ converged before the update of $W$.

%%%%%%%%%%%%%%%%%%%%%%%%%%%%%%%%%%%%%%%%%%%%%%%%%%%%%%%%%%%%%%%%%%%%%%%%%%%%%%%%%%%%%%%%%
\section{Onsager's critical point at finite ${\bf M}$\label{SecOnsager}}
%%%%%%%%%%%%%%%%%%%%%%%%%%%%%%%%%%%%%%%%%%%%%%%%%%%%%%%%%%%%%%%%%%%%%%%%%%%%%%%%%%%%%%%%%

\begin{figure}[t]
\vspace{+0.5cm}
\includegraphics[width=0.99\columnwidth,clip=true]{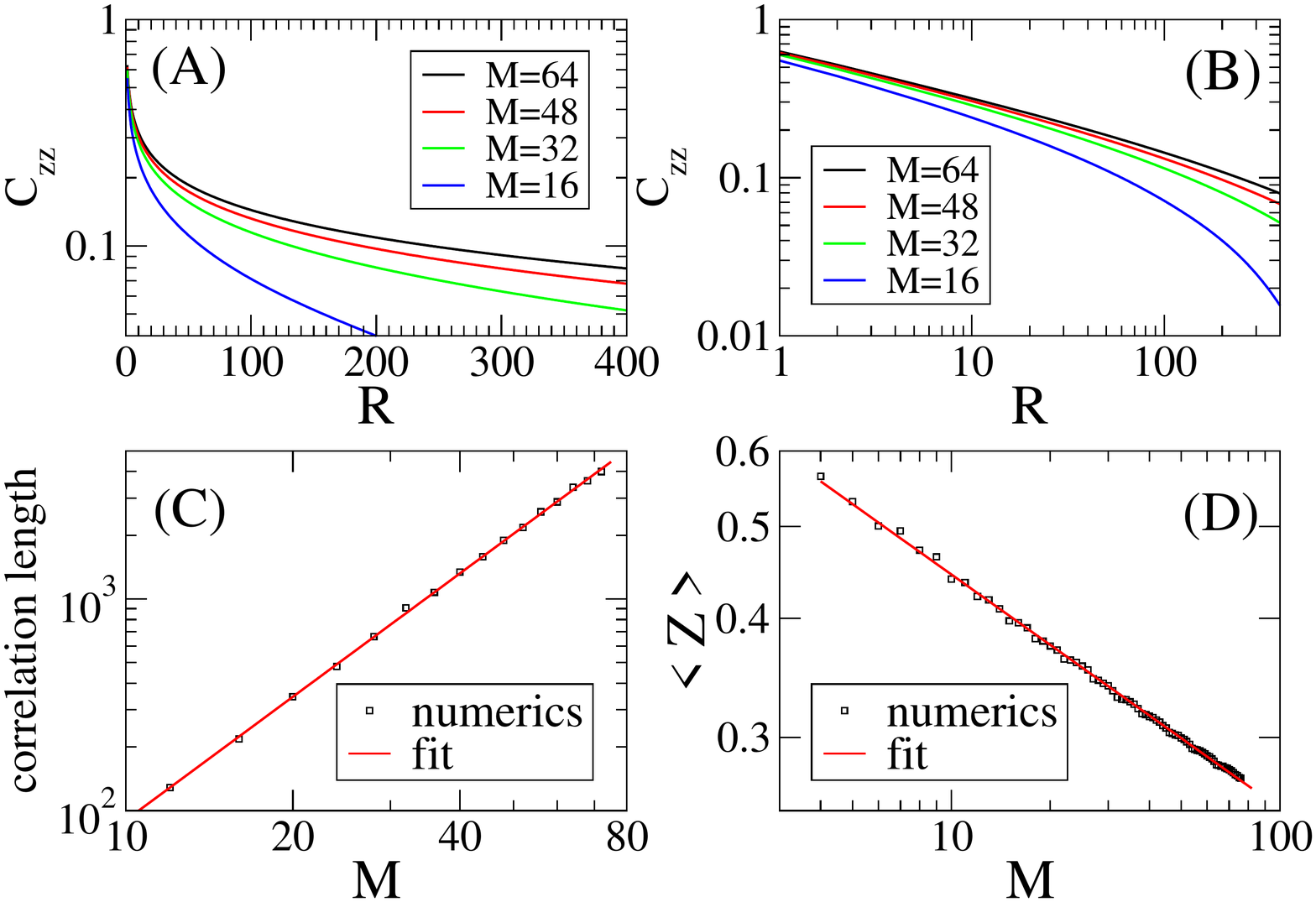}
\vspace{-0.9cm}
\caption{ 
In A and B,
the ferromagnetic correlator, 
$
C_{zz}(R)=
\langle Z_{(x+R,y)} Z_{(x,y)}\rangle-
\langle Z \rangle^2
$,
at the critical $\beta_c=\beta_0$ in the classical Ising model
with zero transverse field and longitudinal bias.
The range of the correlator increases with the environmental bond dimension $M$.
In A,
a logarithmic plot emphasizing exponential tails for large $R$:
$
C_{zz}(R) \sim e^{-R/\xi}.
$
The correlation length $\xi$ is shown in panel C.
In B,
a log-log plot emphasizing a power law decay for intermediate $1\ll R\ll\xi$, 
$
C_{zz}(R) \sim R^{-\eta},
$  
converging to $\eta=1/4$ with increasing $M$.
In C,
the correlation length $\xi$ as a function of $M$. 
The best fit $\xi=1.05M^{1.93}$ suggests a power-law divergence with increasing $M$. 
In D,
spontaneous magnetization $\langle Z\rangle$ as a function of $M$.
The best fit $\langle Z\rangle=0.784M^{-0.246}$ suggests a power-law
decay towards the exact $\langle Z\rangle=0$.
}
\label{FigOnsager}
\end{figure}

In principle, 
a PEPS with a finite bond dimension $D$ can describe a finite-temperature critical point exactly. 
The best example is the Onsager's phase transition at $\beta_0$ in the classical Ising model with $h=0=\delta$, 
where the exact PEPS tensor with $D=2$ is
\bea
A^{ia}_{s_t,s_r,s_b,s_l} = 
T^{ia}_{s_t,s_r,s_b,s_l}(\beta_0),
\label{AOnsager}
\eea
compare Eq. (\ref{T}).
Thus the problem at criticality is not a finite $D$ but a finite environmental bond dimension $M$,
as demonstrated in Fig. \ref{FigOnsager},
where we provide numerical evidence that the correlation length diverges with increasing $M$ roughly like $\xi\sim M^2$ 
and the spontaneous magnetization decays like $\xi^{-1/8}\sim M^{-1/4}$.
Here $1/8$ is the scaling dimension. 
With increasing $M$ the effective environment does converge to the exact one 
but not at an exponential rate.
Apparently,
a finite $M$ cannot provide an accurate effective environment with an infinite correlation length.

For a nonzero transverse field $h$, 
the PEPS tensor at finite $\beta$ is obtained after a series of Suzuki-Trotter steps. 
An accurate step across the critical point requires an accurate effective environment 
that cannot be obtained with a finite $M$.
We bypass this problem by adding a tiny longitudinal bias field 
that makes both the correlation length and the necessary $M$ finite.    

%%%%%%%%%%%%%%%%%%%%%%%%%%%%%%%%%%%%%%%%%%%%%%%%%%%%%%%%%%%%%%%%%%%%%%%%%%%%
%%%%%%%%%%%%%%%%%%%%%%%%%%%%%%%%%%%%%%%%%%%%%%%%%%%%%%%%%%%%%%%%%%%%%%%%%%%%

%%%%%%%%%%%%%%%%%%%%%%%%%%%%%%%%%%%%%%%%%%%%%%%%%%%%%%%%%%%%%%%%%%%%%%%%%%%%%%%%%%%%%%%%%

\begin{thebibliography}{99}



\bibitem{White} S. R. White, 
                Phys. Rev. Lett. {\bf 69}, 2863 (1992).

\bibitem{Schollwoeck} U. Schollw\"ock,
                      Annals of Physics {\bf 326}, 96 (2011).

\bibitem{PEPS} F. Verstraete and J. I. Cirac, cond-mat/0407066; 
               V. Murg, F. Verstraete, and J. I. Cirac, 
               Phys. Rev. A {\bf 75}, 033605 (2007);
               G. Sierra and M. A. Martın-Delgado, 
               arXiv:cond-mat/9811170; 
               T. Nishino and K. Okunishi, 
               J. Phys. Soc. Jpn. {\bf 67}, 3066 (1998); 
               Y. Nishio, N. Maeshima, A. Gendiar, and T. Nishino, 
               cond-mat/0401115;
               J. Jordan, R. Or\'us, G. Vidal, F. Verstraete, and J. I. Cirac,
               Phys. Rev. Lett. {\bf 101}, 250602 (2008);
               Z.-C. Gu, M. Levin, and X.-G. Wen, 
               Phys. Rev. B {\bf 78}, 205116 (2008);
               H. C. Jiang, Z. Y. Weng, and T. Xiang, 
               Phys. Rev. Lett. {\bf 101}, 090603 (2008);
               Z. Y. Xie, H. C. Jiang, Q. N. Chen, Z. Y. Weng, and T. Xiang, 
               Phys. Rev. Lett. {\bf 103}, 160601 (2009);
               P.-C. Chen, C.-Y. Lai, and M.-F. Yang, 
               J. Stat. Mech.: Theory Exp. (2009) P10001;
               R. Or\'us and G. Vidal, 
               Phys. Rev. B {\bf 80}, 094403 (2009).
               
\bibitem{MERA} G. Vidal, 
               Phys. Rev. Lett. {\bf 99}, 220405 (2007); 
               G. Vidal,
               Phys. Rev. Lett. {\bf 101}, 110501 (2008); 
               {\L}. Cincio, J. Dziarmaga, and M. M. Rams, 
               Phys. Rev. Lett. {\bf 100}, 240603 (2008); 
               G. Evenbly and G. Vidal, 
               Phys. Rev. Lett. {\bf 102}, 180406 (2009); 
               G. Evenbly and G. Vidal, 
               Phys. Rev. B {\bf 79}, 144108 (2009).
               
\bibitem{branching} G. Evenbly and G. Vidal, 
                    Phys. Rev. Lett. {\bf 112}, 240502 (2014);
                    Phys. Rev. B {\bf 89}, 235113 (2014) 

\bibitem{fermions}   T. Barthel, C. Pineda, and J. Eisert
                     Phys. Rev. A {\bf 80}, 042333 (2009); 
                     P. Corboz and G. Vidal, 
                     Phys. Rev. B {\bf 80}, 165129 (2009);
                     P. Corboz, G. Evenbly, F. Verstraete, and G. Vidal, 
                     Phys. Rev. A {\bf 81}, 010303(R) (2010);
                     C. V. Kraus, N. Schuch, F. Verstraete, and J. I. Cirac,
                     Phys. Rev. A {\bf 81}, 052338 (2010);
                     C. Pineda, T. Barthel, and J. Eisert, 
                     Phys. Rev. A {\bf 81}, 050303(R) (2010).
                     Z.-C. Gu, F. Verstraete, and X.-G. Wen, 
                     arXiv:1004.2563 (2010).                      
               
\bibitem{highTc} J. Hubbard, 
                 Proc. Roy. Soc. (London), Ser. A {\bf 276}, 238 (1963); 
                 P. W. Anderson, 
                 Science {\bf 235}, 1196 (1987).

\bibitem{PEPStJ}  P. Corboz, R. Or\'us, B. Bauer, and G. ́Vidal,
                  Phys. Rev. B {\bf 81}, 165104 (2010);
                  P. Corboz, S. R. White, G. Vidal, and M. Troyer,
                  Phys. Rev. B {\bf 84}, 041108 (2011);
                  P. Corboz, T. M. Rice, M. Troyer,
                  Phys. Rev. Lett. {\bf 113}, 046402 (2014). 

\bibitem{VMC} D. A. Ivanov. 
              Phys. Rev. B {\bf 70}, 104503 (2004);
              W.-J. Hu, F. Becca, S. Sorella,
              Phys. Rev. B {\bf 85}, 081110(R) (2012).
              
\bibitem{WhiteKagome} S. Yan, D. A. Huse, and S. R. White,
                      Science {\bf 332}, 1173 (2011). 

\bibitem{CincioVidal} L. Cincio and G. Vidal, 
                      Phys. Rev. Lett. {\bf 110}, 067208 (2013).

\bibitem{PepsRVB} D. Poilblanc, N. Schuch, D. P\'{e}rez-Garc\'{i}a, and J. I. Cirac,
                  Phys. Rev. B {\bf 86}, 014404 (2012). 

\bibitem{PepsKagome} D. Poilblanc, N. Schuch,
                     Phys. Rev. B {\bf 87}, 140407(R) (2013).
                     
\bibitem{PepsJ1J2} L. Wang, D. Poilblanc, Z.-C. Gu, X.-G. Wen, and F. Verstraete,
                   Phys.Rev.Lett. {\bf 111}, 037202 (2013).              
                                          
\bibitem{ancillas} F. Verstraete, J. J. Garcia-Ripoll, and J. I. Cirac,
                   Phys. Rev. Lett. {\bf 93}, 207204 (2004); 
                   M. Zwolak and G. Vidal, 
                   Phys. Rev. Lett. {\bf 93}, 207205 (2004);
                   A.E. Feiguin and S.R. White, 
                   Phys. Rev. B {\bf 72}, 220401 (2005).
                                             
\bibitem{WhiteT} S. R. White, 
                 arXiv:0902.4475;
                 E.M. Stoudenmire and Steven R. White,
                 New J. Phys. {\bf 12}, 055026 (2010);
                 I. Pizorn, V. Eisler, S. Andergassen, and M. Troyer,
                 New J. Phys. {\bf 16}, 073007 (2014).    
                 
\bibitem{Czarnik} P. Czarnik, {\L}. Cincio, and J. Dziarmaga, 
                  Phys. Rev. B {\bf 86}, 245101 (2012);
                  P. Czarnik and J. Dziarmaga,
                  Phys. Rev. B {\bf 90}, 035144 (2014).         
                     
\bibitem{ChinaT} Z. Y. Xie, H. C. Jiang, Q. N. Chen, Z. Y. Weng, T. Xiang,
                 Phys.Rev.Lett. {\bf 103}, 160601 (2009);
                 H.H. Zhao, Z.Y. Xie, Q.N. Chen, Z.C. Wei, J.W. Cai, T. Xiang,
                 Phys. Rev. B {\bf 81}, 174411 (2010);   
                 W. Li, S.-J. Ran, S.-S. Gong, Y. Zhao, B. Xi, F. Ye, and G. Su, 
                 Phys. Rev. Lett. {\bf 106}, 127202 (2011);
                 Z. Y. Xie, J. Chen, M. P. Qin, J. W. Zhu, L. P. Yang, and T. Xiang, 
                 Phys. Rev. B {\bf 86}, 045139 (2012);
                 Shi-Ju Ran, Wei Li, Bin Xi, Zhe Zhang and Gang Su,
                 Phys. Rev. B {\bf 86}, 134429 (2012);
                 S.-J. Ran, B. Xi, T. Liu, and G. Su, 
                 Phys. Rev. B {\bf 88}, 064407 (2013);
                 A. Denbleyker, Y. Liu, Y. Meurice, M. P. Qin, T. Xiang, Z. Y. Xie, J. F.     Yu, H. Zou,
                 Phys. Rev. D {\bf 89}, 016008 (2014). 
                               
\bibitem{Poulin} D. Poulin and M. B. Hastings, 
                 Phys. Rev. Lett. {\bf 106}, 080403 (2011); 
                 A. J. Ferris and D. Poulin, 
                 Phys. Rev. B {\bf 87}, 205126 (2013).                
                                                                     
\bibitem{CMR} R. J. Baxter, 
              J. Math. Phys. {\bf 9}, 650 (1968); 
              J. Stat. Phys. {\bf 19}, 461 (1978); 
              T. Nishino and K. Okunishi, 
              J. Phys. Soc. Jpn. {\bf 65}, 891 (1996);
              R. Or\'us and G. Vidal, 
              Phys. Rev. B {\bf 80}, 094403 (2009);
              R. Or\'us, 
              Phys. Rev. B {\bf 85}, 205117 (2012);
              R. Or\'us, 
              Ann. of Phys. {\bf 349}, 117 (2014);
              Ho N. Phien, I. P. McCulloch, G. Vidal,
              arXiv:1411.0391.

\bibitem{hc} H. Rieger, N. Kawashima, 
             Europ. Phys. J. B {\bf 9}, 233 (1999);
             H.W.J. Blote and Y. Deng, 
             Phys. Rev. E {\bf 66}, 066110 (2002).
             
\bibitem{OPS} F. Verstraete, J.I. Cirac, V. Murg,
              Adv. Phys. {\bf 57}, 143 (2008).
      

\end{thebibliography}
\end{document}